\documentclass[twocolumn]{aastex6} % Just fine

\newcommand{\ind}[1]{_{\mathrm{#1}}}

\newcommand{\Dnu}{\Delta\nu}

\shorttitle{Testing the asteroseismic scaling relations}
\shortauthors{Gaulme et al.}

\begin{document}

%% LaTeX will automatically break titles if they run longer than
%% one line. However, you may use \\ to force a line break if
%% you desire.

\title{Testing the asteroseismic scaling relations for Red Giants with Eclipsing Binaries Observed by Kepler}

%% Use \author, \affil, and the \and command to format
%% author and affiliation information.
%% Note that \email has replaced the old \authoremail command
%% from AASTeX v4.0. You can use \email to mark an email address
%% anywhere in the paper, not just in the front matter.
%% As in the title, use \\ to force line breaks.
\author{P. Gaulme\altaffilmark{1,2}}
\author{J. McKeever\altaffilmark{1}}
\author{J. Jackiewicz\altaffilmark{1}}
\author{M. L. Rawls\altaffilmark{1}}
\author{E. Corsaro\altaffilmark{3,4,5}}
\author{B. Mosser\altaffilmark{6}}
\author{J. Southworth\altaffilmark{7}}
\author{S. Mahadevan\altaffilmark{8,9}}
\author{C. Bender\altaffilmark{8,9}}
\author{R. Deshpande\altaffilmark{8,9}}

\email{gaulme@nmsu.edu}

\altaffiltext{1}{Department of Astronomy, New Mexico State University, P.O. Box 30001, MSC 4500, Las Cruces, NM 88003-8001, USA}
\altaffiltext{2}{Apache Point Observatory, 2001 Apache Point Road, P.O. Box 59, Sunspot, NM 88349, USA}
\altaffiltext{3}{Laboratoire AIM, CEA/DRF-CNRS, Universit\'e Paris 7 Diderot, IRFU/SAp, Centre de Saclay, 91191 Gif-sur-Yvette, France}
\altaffiltext{4}{Instituto de Astrof\'{\i}sica de Canarias, E-38200 La Laguna, Tenerife, Spain}
\altaffiltext{5}{Departamento de Astrof\'{\i}sica, Universidad de La Laguna, E-38205 La Laguna, Tenerife, Spain}
\altaffiltext{6}{LESIA, Observatoire de Paris, PSL Research University, CNRS, Universit\'e Pierre et Marie Curie,Universit\'e Denis Diderot, 92195 Meudon, France}
\altaffiltext{7}{Astrophysics Group, Keele University, Staffordshire, ST5 5BG, UK}
\altaffiltext{8}{Department of Astronomy \& Astrophysics, The Pennsylvania State University, 525 Davey Lab, University Park, PA-16802}
\altaffiltext{9}{Center for Exoplanets $\&$ Habitable Worlds, The Pennsylvania State University, 525 Davey Lab, University Park, PA-16802}

%% Notice that each of these authors has alternate affiliations, which
%% are identified by the \altaffilmark after each name.  Specify alternate
%% affiliation information with \altaffiltext, with one command per each
%% affiliation.

%\altaffilmark{1}

%% Mark off your abstract in the ``abstract'' environment. In the manuscript
%% style, abstract will output a Received/Accepted line after the
%% title and affiliation information. No date will appear since the author
%% does not have this information. The dates will be filled in by the
%% editorial office after submission

\begin{abstract}
Given the potential of ensemble asteroseismology for understanding fundamental properties of large numbers of stars, it is critical to determine the accuracy of the scaling relations on which these measurements are based. From several powerful validation techniques, all indications so far show that stellar radius estimates from the asteroseismic scaling relations are accurate to within a few percent. Eclipsing binary systems hosting at least one star with detectable solar-like oscillations constitute the ideal test objects for validating asteroseismic radius and mass inferences. By combining radial-velocity measurements and photometric time series of eclipses, it is possible to determine the masses and radii of each component of a double-lined spectroscopic binary. We report the results of a four-year radial-velocity survey performed with the \'echelle spectrometer of the Astrophysical Research Consortium's 3.5-m telescope and the APOGEE spectrometer at Apache Point Observatory. We compare the masses and radii of 10 red giants obtained by combining radial velocities and eclipse photometry with the estimates from the asteroseismic scaling relations. We find that the asteroseismic scaling relations overestimate red-giant radii by about 5\,\% on average and masses by about 15\,\% for stars at various stages of red-giant evolution. Systematic overestimation of mass leads to underestimation of stellar age, which can have important implications for ensemble asteroseismology used for Galactic studies. As part of a second objective, where asteroseismology is used for understanding binary systems, we confirm that oscillations of red giants in close binaries can be suppressed enough to be undetectable, an hypothesis that was proposed in a previous work.
\end{abstract}

%% Keywords should appear after the \end{abstract} command. The uncommented
%% example has been keyed in ApJ style. See the instructions to authors
%% for the journal to which you are submitting your paper to determine
%% what keyword punctuation is appropriate.


%%%%%%%%%%%%%%%%%%%%%%%%%%%%%%%%%%%%%%%%%%%%%%%%%%%%%%%%%%%%%%%
%                                                                                  INTRODUCTION
%%%%%%%%%%%%%%%%%%%%%%%%%%%%%%%%%%%%%%%%%%%%%%%%%%%%%%%%%%%%%%%
\section{Oscillating Red Giants in Eclipsing Binaries}
\label{sect_update}
The simplest analysis of asteroseismic data is based on the overall properties of the oscillations, which are the frequency of their maximum amplitude $\nu\ind{max}$, and the mean frequency separation $\Delta\nu$ between consecutive modes of same degree. Thanks to the pair of asteroseismic scaling relations and a measurement of effective temperature $T\ind{eff}$, one gets an estimate of a star's surface gravity $\log g$ and mean density $\bar{\rho}$ by respectively comparing $\nu\ind{max}$ and $\Delta\nu$  with those of the Sun \citep[e.g.][]{Kjeldsen_Bedding_1995}: 
\begin{eqnarray}
\frac{\bar{\rho}}{\bar{\rho}_\odot}\ &=&\ \left(\frac{\Delta\nu}{\Delta\nu_\odot}\right)^2 \\
\frac{g}{g_\odot}\ &=&\  \frac{\nu\ind{max}}{\nu\ind{max_\odot}}\ \left(\frac{T\ind{eff}}{T\ind{eff,\odot}}\right)^{\frac{1}{2}}.
\end{eqnarray}\\
It is then straightforward to deduce a star's mass $M$ and radius $R$ relatively to the Sun:

\begin{eqnarray}
\frac{R}{R_\odot}\ &=&\ \left(\frac{\nu\ind{max}}{\nu\ind{max_\odot}}\right)\ \ \left(\frac{\Delta\nu_\odot}{\Delta\nu}\right)^2\ \left(\frac{T\ind{eff}}{T\ind{eff,\odot}}\right)^{\frac{1}{2}}\\
\frac{M}{M_\odot}\ &=&\  \left(\frac{\nu\ind{max}}{\nu\ind{max_\odot}}\right)^3\ \left(\frac{\Delta\nu_\odot}{\Delta\nu}\right)^4\ \left(\frac{T\ind{eff}}{T\ind{eff,\odot}}\right)^{\frac{3}{2}}.
\end{eqnarray}\\
In practice, the measurement of the asteroseismic global parameters $\nu\ind{max}$ and $\Delta\nu$ has been largely used to estimate masses and radii of the stars displaying solar like oscillations from the CoRoT and \textit{Kepler} data \citep[see][for recent reviews]{Chaplin_Miglio_2013, Belkacem_2013}.

Given the importance of asteroseismology and its scaling laws, much effort has been carried out to test their validity. We may distinguish two kinds of approaches: those based on validating the relation between $\Delta\nu$ and mean density $\bar{\rho}$ from models and simulated data \citep[e.g.][]{Stello_2009b,White_2011,Miglio_2013}, the others based on measuring $R$ of actual stars independently from asteroseismology \citep[e.g.][]{Huber_2011,Huber_2012,Silva_Aguirre_2012,Baines_2014}.
All works indicated that radius estimates from asteroseismic scaling relations are accurate to a few percent. On the contrary, similar tests with independent mass determination of oscillating stars for individual stars have not been possible so far. Indeed, theoretical studies focused on the reliability of the $\Delta\nu$-$\bar\rho$ scaling relation and not on $\nu\ind{max}$. This is because $\nu\ind{max}$ has no secure theoretical basis, as it is not yet possible to make reliable predictions of the amplitude of stochastically excited modes and their dependence with frequency \citep{Belkacem_2011,Christensen-Dalsgaard_2012}. Observationally, there is some evidence to support the conclusion that the scaling relations do provide biased masses in some instances. Epstein et al. (2014, ApJ 785, 28) have found that the masses of metal-poor halo giants are significantly overestimated. White et al. (2013, MNRAS 433, 1262) found that combining the interferometric radii with the asteroseismic density implied a mass for the F star $\theta$ Cyg that was significantly lower than expected from its position in the Hertzsprung-Russell diagram.

Eclipsing binaries systems (EBs) hosting at least one star with detectable solar-like oscillations constitute an ideal test case. Indeed, it is possible to determine the projected masses of each component ($M_1 \sin i$, $M_2 \sin i$) for double-line spectroscopic binaries (SB2), and the mass function $M_2^3/(M_1+M_2)^2 \sin^3 i$ for single-line spectroscopic binaries (SB1), where $i$ is the inclination of the orbital plane. For EBs, the inclination $i$ is easily retrieved from modeling the eclipses in the light curves. Absolute stellar radii $(R_1, R_2)$ are obtained from combining radial velocity and eclipse photometric measurements.

So far, all published stars known to both display solar-like oscillations and belong to EBs are red giants (RGs), and all have been detected by the \textit{Kepler} mission. The first detection was the 408-day period system KIC 8410637 \citep{Hekker_2010, Frandsen_2013}. Since then, \citet{Gaulme_2013, Gaulme_2014} reported a list 18 RG eclipsing-binary (RG/EB) candidates, of which 14 displayed oscillations. \citet{Beck_2014,Beck_2015} reported the discovery of 17 stars with tidally-excited pulsations (``heartbeat''), where each system has a RG component with oscillations, and two are also EBs. Two RG/EB systems, KIC 8410637 and 9246715 have been completely characterized in terms of masses and radii by combining photometry and radial velocities \citep{Frandsen_2013, Helminiak_2015,Rawls_2016}. Both show a fairly good agreement between asteroseismic and dynamical estimates of surface gravities and mean densities, even though \citet{Huber_2014} and \citet{Brogaard_2016} contested the agreement regarding KIC 8410637.

\citet{Gaulme_2014} observed that among the 19 RG/EBs identified at the time, four systems did not display oscillations.  This is observed in the closest systems where rotational and orbital periods are almost synchronized and where strong surface activity is detected. They suggested that tidal forces, which tend to synchronize and circularize binary systems, spin up RGs, with this phenomenon becoming stronger as systems are closer. This would lead to the development of a dynamo mechanism, and thus the generation of magnetic fields in the RGs that become visible at the surface. The resulting spots likely absorb part of the pressure mode energy making oscillations impossible to detect in the closest systems. Alternatively, it is proposed that the presence of spots shows that the convective energy is diverted into activity signal and not into global oscillations. This would mean that properties of convection are considerably affected by binarity in the closest systems, and that oscillation excitation is reduced, or suppressed altogether.

In this paper, we report the result of a four-year radial-velocity (RV) survey performed with the \'echelle spectrometer of the Astrophysical Research Consortium (ARC) 3.5-m telescope at Apache Point Observatory (APO). We benefited from complementary observations by the Apache Point Observatory Galactic Evolution Experiment  (APOGEE) spectrograh for one system. The targets are 17 EB systems of the 18 \citet{Gaulme_2013,Gaulme_2014}'s the RG/EB candidates, whose orbital periods range from 15 to 1058 days. Among those, solar-like oscillations are detected in 13, of which nine are SB2s and four SB1s. The remaining four are RG/EB candidates where no oscillations are detected. Our first objective is to test the nature of the 17 systems, where RVs allow us to determine whether the RGs belong to or are aligned with EBs. The second consists of measuring the masses and radii of the four RG candidates with no oscillations to determine if their expected $\nu\ind{max}$ fall in the observable range, i.e.\ not much larger than the Nyquist frequency. The third and main objective is the comparison of masses, radii, mean densities, and surface gravities with those obtained with the asteroseismic scaling relations. For the latter, we consider the nine SB2 with oscillations as well as KIC 8410637 for which we re-estimate its asteroseismic parameters and use \citet{Frandsen_2013}'s dynamical measurement of the mass and radius of each star.

%%%%%%%%%%%%%%%%%%%%%%%%%%%%%%%%%%%%%%%%%%%%%%%%%%%%%%%%%%%%%%%
%                                                                         DATA  AND ANALYSIS
%%%%%%%%%%%%%%%%%%%%%%%%%%%%%%%%%%%%%%%%%%%%%%%%%%%%%%%%%%%%%%%
\section{Data and Analysis}

%===============================================================================================
\subsection{Kepler photometric light curves}
Light curves are used for a double purpose - eclipse and asteroseismic modeling - which entails modeling eclipse shapes by removing stellar activity, and measuring solar-like oscillations by removing eclipse features. The methods we use are described in \citet{Gaulme_2013,Gaulme_2014} and here we provide a summary.
The way light curves are processed is of prime importance as part of the conclusions of this paper depends on our ability to provide reliable estimates of stellar radii. As we indicate in Sect.~\ref{sec_rv}, radii measurements are a function of the relative depths of the eclipses, unlike mass, which is not sensitive to eclipse photometry. We thus need the relative photometry to be calibrated as finely as possible.%This is not a straightforward process, but even though no perfect solution can be achieved, the goal of providing reliable radii is met.

We work with the raw SAP\_FLUX measurements, which are the fluxes integrated per mask aperture, available on the MAST website\footnote{\url{http://archive.stsci.edu/kepler/}}. The major challenge in concatenating light curves and studying stellar activity on periods longer than a quarter is to ensure photometric continuity before and after each interruption. The main cause of photometric jumps from quarter to quarter is the fact that the \textit{Kepler} telescope rotates four times a year, which implies that a given star falls on four different chips. However, the pointing is fine enough that a star repeatedly covers the same group of pixels every four quarter. Light curves are obtained by adding the pixels of the masks that are designed for every star of the field of view. For  a given star, a mask is designed for each of \textit{Kepler}'s positions. Because of the photo response non-uniformity (PRNU) of the pixels and the changing size of the masks, the recorded flux changes. Both PRNU and varying mask areas lead to flux discontinuity that should be adjusted in a multiplicative way. The first correction we apply is therefore a normalization that turns the photoelectric counts into relative flux, by dividing each quarter's light curve by its average. A median is actually more appropriate than a mean as outliers and large photometric jumps can bias the mean. If photometric variations would only be generated by PRNU and masks, this process should be enough. As a matter of fact, this is true for systems where no stellar activity is measurable, if we exclude the effect of the differential velocity aberration (see below).

Issues arise with the systems that display strong pseudo-periodic luminosity fluctuations. For those, the average (or median) over a quarter is biased by the fact that the number of pseudo periods is too small to be averaged out. In our cases, pseudo periods range from 15 to about 60 days. Therefore, the median is not a perfect estimator of the mean photometry. This is an intrinsic limitation of the light curve photometric accuracy. In such cases, jumps remain, of amplitudes within a few percent. Given that the remaining jumps are caused by a biased normalization, the second layer of adjustment to be applied should still be done in a multiplicative way. However, this is not possible in practice because none of the quarters can be considered as an absolute reference. The only corrections we may apply are additive, to ensure a smooth aspect of the light curve and to minimize their effects in the Fourier domain. This explains why residuals are larger when modeling systems with large photometric activity (KIC 7943602, 3955867, 4569590, 9291629, see Fig.~\ref{fig_ph_rv_sb2}). When modeling such systems, what matters is that the residuals are symmetrically distributed around the best model light-curve. Note that this discussion regards mostly the systems with no oscillations, and does not have any significant impact on the systems used to test the asteroseismic scaling laws.

We employ two ways to smooth the remaining discontinuities once quarters are divided by their median. When a gap is short with respect to the photometric variability timescale, each side of the gap is adjusted accordingly. When a gap is longer than the variability period, we simply adjust the photometry with the difference of the means of each chunk surrounding the gap. Once the complete time series is leveled and concatenated, a linear fit is subtracted from it to take into account the decreasing instrumental sensitivity. Finally, we compensate for the differential velocity aberration -- the motion of the target across a fixed aperture smaller than the point spread function -- caused by the pixel scale breathing along the satellite's orbit (372.5 days), whose peak-to-peak amplitude ranges from 0.5\,\% to 3.8\,\%. This is done by subtracting from each light curve a 372.5-day period sine fitting and a first harmonic, which is enough to reduce its amplitude to less than 0.5\,\%.

For asteroseismic analysis, we remove the data corresponding to the eclipses and bridge them with a second-order polynomial, constrained by the surrounding data. We then smooth the eclipse-less light curve on a large number of points (about 1000) and subtract it from the original clean light curve to get a flattened time series, which we use for eclipse modeling.

For asteroseismology we work with the power density spectra of the light curves.  To minimize the effects of the incomplete duty cycle, we perform gap fillings and make use of the fast Fourier transform (FFT).  All short gaps (only several missing points) are interpolated with a second order polynomial estimated from the nearby data points. Long gaps are filled with zeros. To reduce the impact of abrupt discontinuities around long gaps, the edges of each section in between gaps are apodized with a cosine function. This is particularly important when significant variability is detected.  Overall, the duty cycle for these objects is always greater than 85\,\%.

%-----------------------------------------------------
\begin{figure*}[t]
\epsscale{1}
\plotone{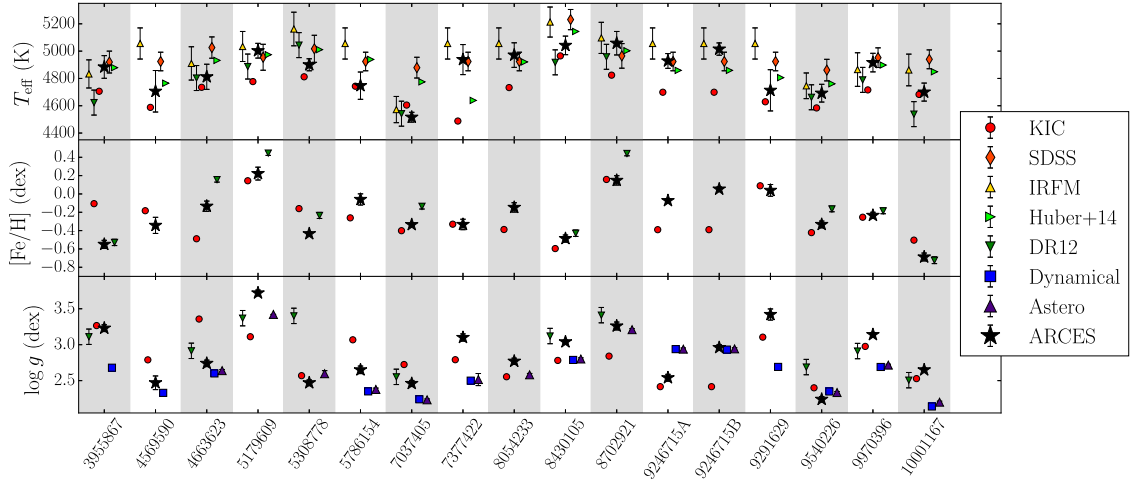}
\caption{Comparisons of $T\ind{eff}$ (top), [Fe/H] (middle) and $\log g$ (bottom) from various sources to those derived by the spectroscopic analysis in this work (black stars). Photometrically derived values include: Kepler Input Catalog as red circles \citep{2011AJ....142..112B}, SDSS $griz$ filter method as red diamonds and IRFM values as yellow triangles (up) \citep{2012ApJS..199...30P}, revised KIC values as light green triangles (right) \citep{2014ApJS..211....2H}. IR spectroscopic values from DR12 of APOGEE are shown as green triangles (down) \citep{2015ApJS..219...12A}.  Also indicated are the dynamically derived (blue squares) and asteroseismic (purple triangles) gravities.
\label{fig_atm}}
\end{figure*}
%-----------------------------------------------------

%===============================================================================================
\subsection{Spectroscopy and stellar parameters}
All spectra were obtained with the \'echelle spectrometer of the 3.5-m ARC telescope at APO (ARCES), except a set of 25 spectra of the system KIC 7037405 from the APOGEE spectrometer of the Sloan Digital Sky Survey \citep[SDSS,][]{Eisenstein_2011}, also based at APO. Details on how the RVs were determined from the ARC \'echelle spectra are described in \citet{Rawls_2016}. 
As regards APOGEE,  we followed the procedures described in \citet{Bender_2012} and Mahadevan et al. (in prep).  In brief, we start with the extracted and wavelength calibrated ``visit'' spectra produced by the APOGEE data reduction pipeline \citep{Nidever_2015}. These spectra are further cleaned of processing residuals resulting from incomplete corrections by the pipeline of telluric absorption and OH emission features, and continuum normalized.  We do not utilize the pipeline derived RVs, but instead use one-dimensional and two-dimensional cross-correlation analysis against template spectra, through a customized implementation of the TODCOR algorithm \citep{Zucker_Mazeh_1994}. Templates are constructed from the PHOENIX based BT-Settl model \citep{Allard_2011} corresponding to the APOGEE derived $T\ind{eff}$, $\log g$, and [M/H], convolved to the APOGEE spectral resolution of 22,500, and rotationally broadened using a four parameter non-linear limb-darkening model \citep{Claret_2012}. Radial velocity measurements are provided in the Appendix.

To determine stellar atmospheric parameters from ARCES data, individual spectra for each object were co-added after removing the derived RV shift due to the motion of the RG at each epoch. The resulting composite spectrum has a higher signal-to-noise ratio (SNR) and ignores the contribution of the companion star, whose flux is typically a few percent of the total. While this should certainly remove the spectral lines of the companion (which has a different radial velocity than the RG), there should still be a contribution from the continuum. By assuming all lines are equally damped -- i.e., the effect of the continuum is felt equally across the entire spectrum --  then the net effect is a shift in the abundance. Thus, a dilution of lines (smaller measured EW) leads to lower abundance estimation.

We used the MOOG spectral synthesis code \citep{2012ascl.soft02009S} to derive the spectroscopic parameters $T_{\rm eff}$, $\log g$, [Fe/H], and microturbulence of the RG star. The technique assumes LTE to achieve ionization and excitation balance for iron lines in the stellar spectrum. This is done using the equivalent
width (EW) of the iron lines and an appropriate atmosphere model. In our case we used a grid of Kurucz ATLAS9 plane-parallel model
atmospheres \citep{2004astro.ph..5087C}. A review of the process can be found in \citet{2014arXiv1407.5817S}. We use a set of 120 FeI lines and 17 FeII lines optimized for cool stars \citep{2013A&A...555A.150T}. The EWs in our spectra were measured using the automated EW finder
ARES \citep{2007A&A...469..783S,2015A&A...577A..67S}. A visual inspection of the output from ARES was performed to ensure that only clear and easily visible lines were included in the rest of the analysis.  We follow the algorithm outlined in \citet{2013A&A...558A..38M} to quickly
arrive at the best solution and associated errors.
%[[[[VSINI?]]]] --> During revision process

\begin{figure*}[t]
\epsscale{0.37}
\plotone{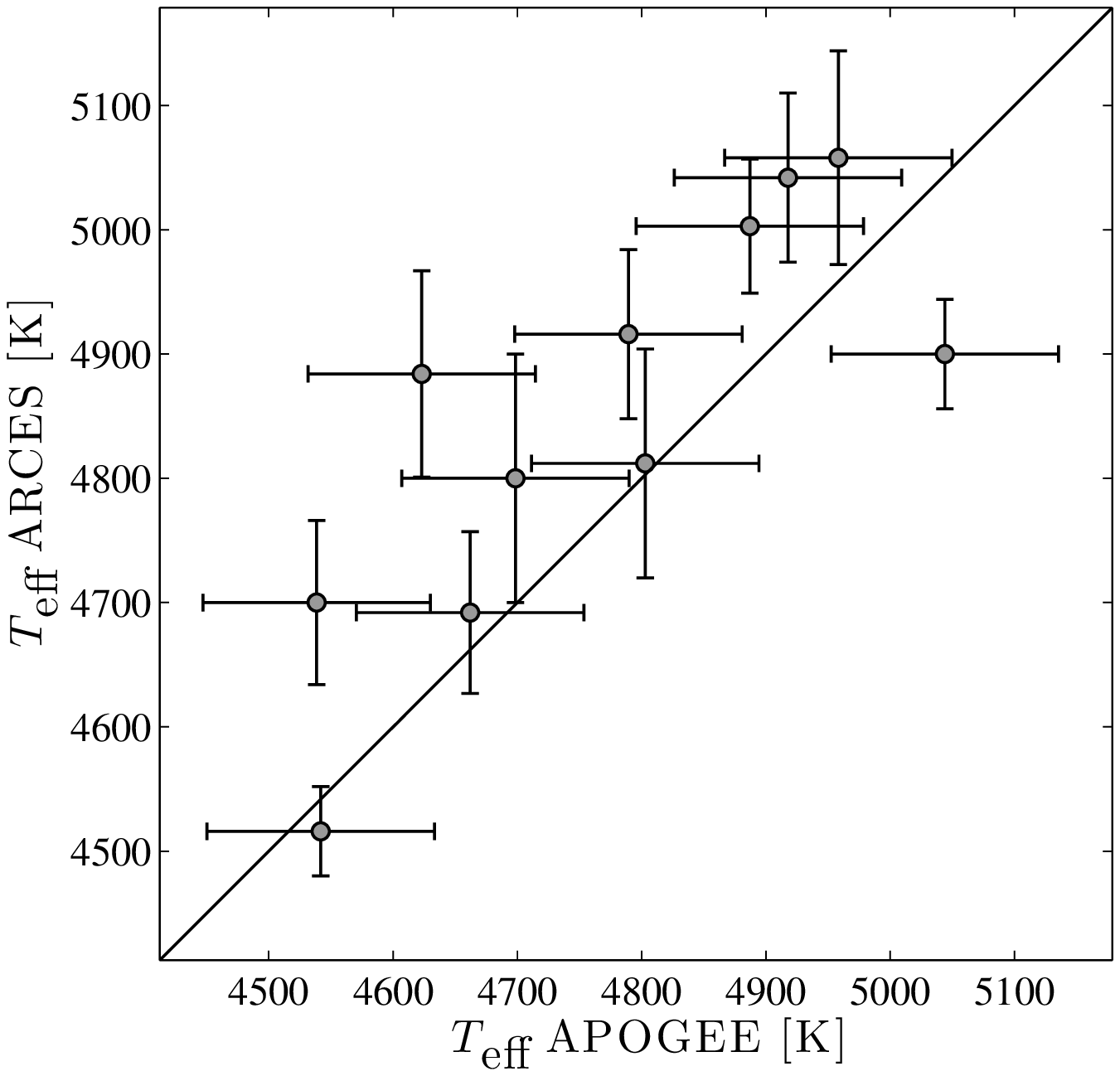}
\plotone{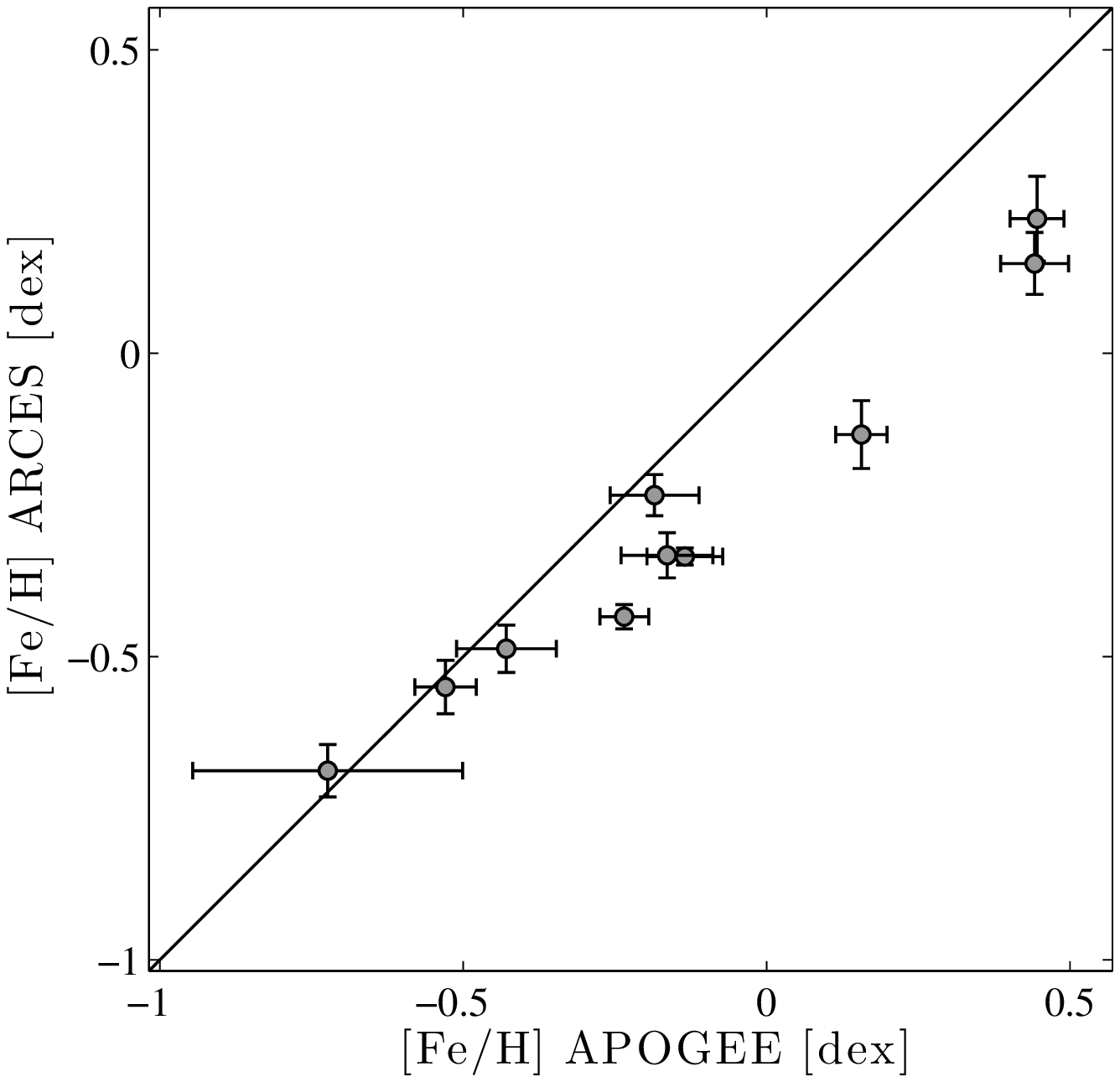}
\plotone{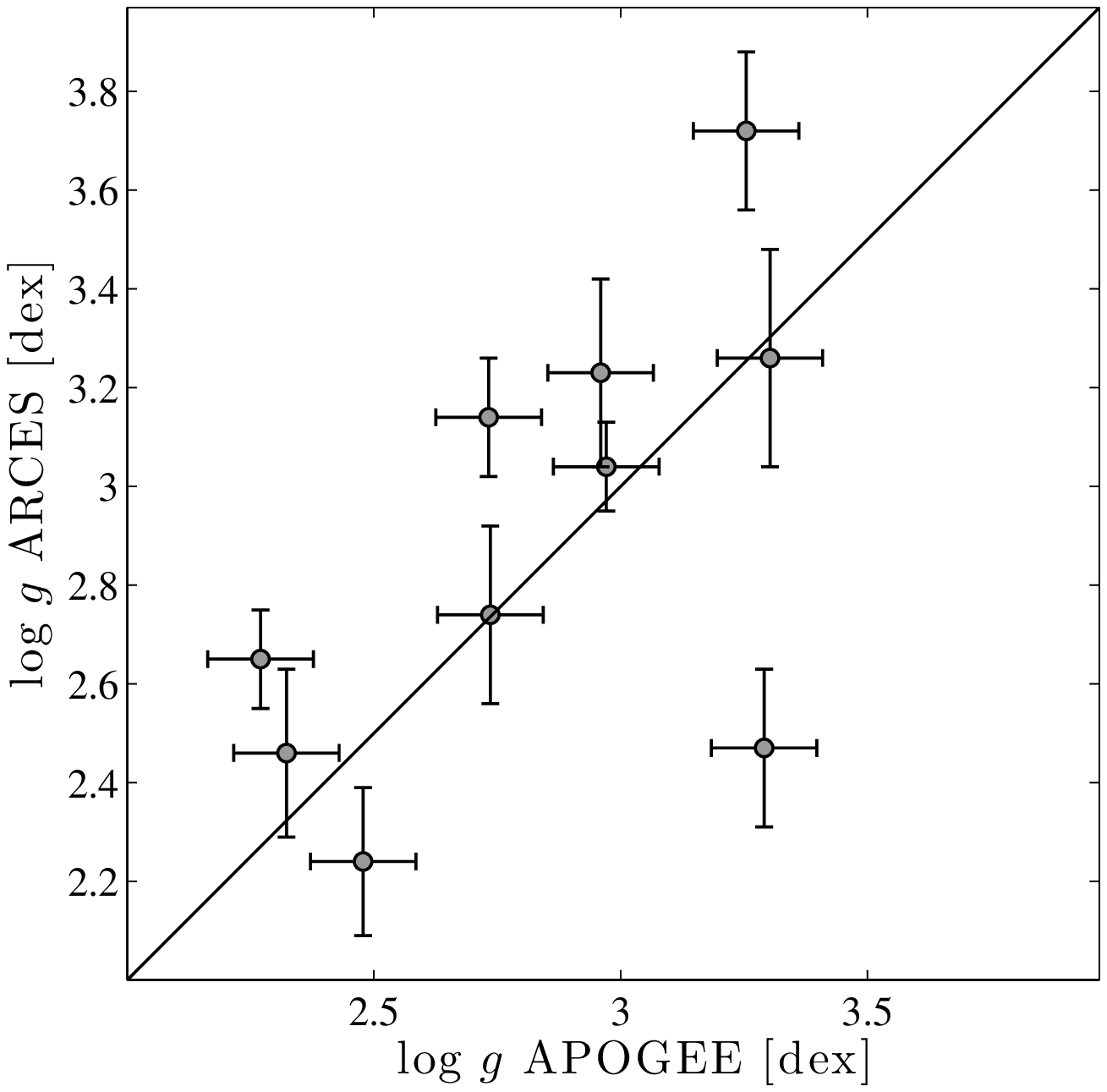}
  \caption{Comparison of the red giant atmospheric parameters ($T\ind{eff}$, [Fe/H], $\log g$) obtained with ARCES (visible)  and APOGEE (IR) measurements. \label{fig_arces_apogee}}
\end{figure*}

The derived spectroscopic parameters are provided in Table~\ref{tab_atm}. Figure~\ref{fig_atm} compares these values with other published results from various catalogs, such as the Kepler Input Catalog and the APOKASC catalog. In Fig.~\ref{fig_arces_apogee}, we specifically compare the measurements using visible spectra and APOGEE infrared spectra. The visible spectra have systematically larger $T_{\rm eff}$ values, as well as lower metallicities, with the discrepancy increasing for more metal-rich stars. The $\log g$ values show better agreement within the uncertainties. We choose to use only the atmospheric parameters we retrieve from the ARCES visible spectra, instead of APOGEE's, for two reasons. Firstly, APOGEE spectra are available for only about half of the systems, and we prefer working with a consistent set of temperatures obtained from the same instrument and data processing routines. Secondly, even though APOGEE data are less sensitive to the companion's line, the data are processed in a massive automatic  pipeline, while we worked with the ARCES spectra one-by-one. To confirm that our results are of good quality, we tested the method on well-known red giants that we also observed with ARCES to calibrate our method. Finally, when comparing our temperature estimates with those available in the literature (Fig.~\ref{fig_atm}), our measurements are very consistent with the average of what was obtained independently. 
%[[INTERPRET LATER???]] --> Not a deep spectroscopic paper. Not sure. We should just show the agreement/disagreement without trying to understand it. I think we're not able.

%-----------------------------------------------------
%-----------------------------------------------------

%===============================================================================================
\subsection{Physical properties from eclipses and radial velocities}\label{sec_rv}
Eclipse modeling consists of retrieving the physical parameters of a binary system from the eclipse duration, depth and shape.  These systems are composed of an RG and a companion that is usually a main-sequence star. In this paper, as well as in \citet{Gaulme_2013,Gaulme_2014}, we make the sacrilegious choice of defining the primary eclipse to be that where the companion star passes in front of the RG. The reason is that all of our systems but KIC 9246715 \citep[discussed in the present work but analyzed in][]{Rawls_2016} are composed of a small companion and a red giant, which makes the exoplanet terminology easier to use: primary or secondary refer to the radius instead of the temperature ratio. Thus, for cases composed of a sun-like main-sequence star and a cooler but larger RG, secondary eclipses are deeper than primary ones. Regarding their shapes, primary eclipses are dominated by the RG limb-darkening function, whereas secondary eclipses are flat except during ingress and egress.

For a given system, we simultaneously fit the \textit{Kepler} light curve and the RVs with the help of the JKTEBOP software \citep{Southworth_2013}. In the case of an SB2, JKTEBOP allows the retrieval of a system's orbital parameters (period, time and argument of periastron, eccentricity, inclination, semi-major axis) and stellar physical properties (masses, radii, temperature ratio). Semi-major axes, radii, masse are deduced from a set of fitted parameters, which are the ratio of radii, central surface brightness ratio, sum of relative radii, limb darkening, inclination, amount of contamination by third light, time of one of the eclipses, semi-amplitude and offset of RVs. We fit the eccentricity $e$ and argument of periastron $\omega$ through the set of parameters $e\cos\omega$ and $e\sin\omega$, which are less degenerate and provide more stable solutions. Since we work only with single-band photometry (\textit{Kepler}'s), we have no robust constraint on the contamination factor, thus we fix it at the mean value provided by the KIC (average over the four orientations of the satellite). All fitted parameters are displayed in Table~\ref{tab_orb}.

%-----------------------------------------------------
\begin{figure*}[h!]
\epsscale{0.33}
\plotone{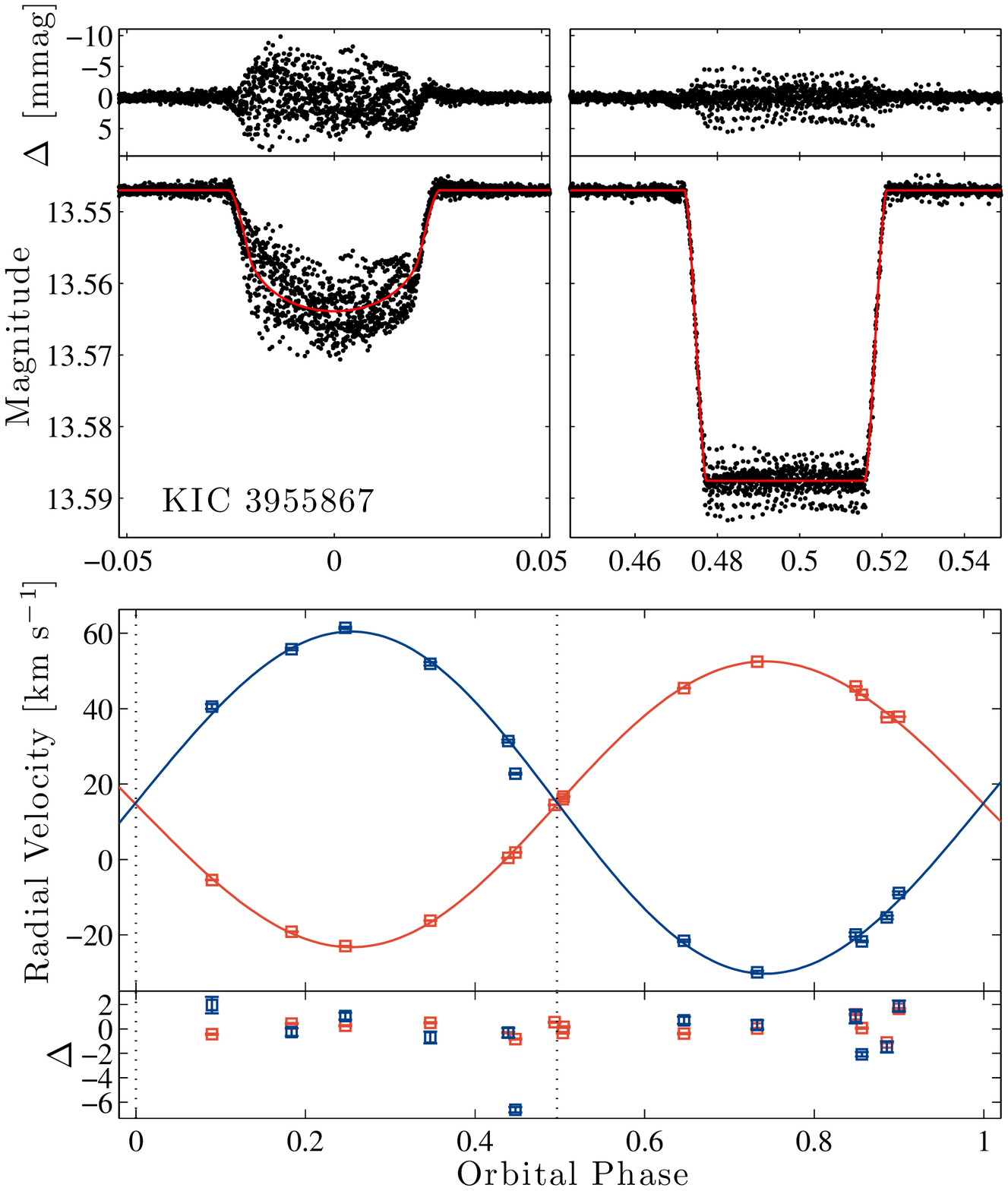}
\plotone{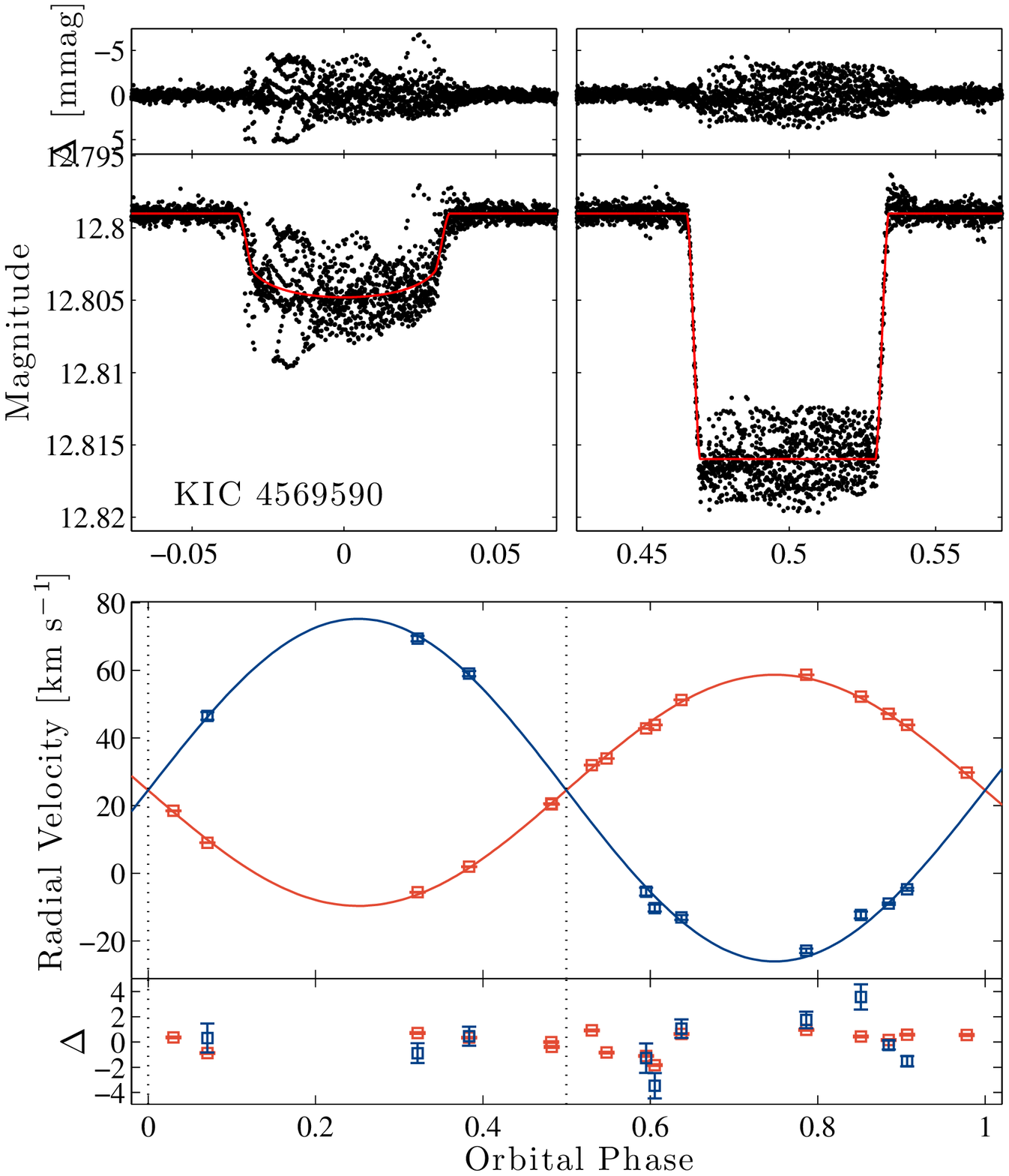}
\plotone{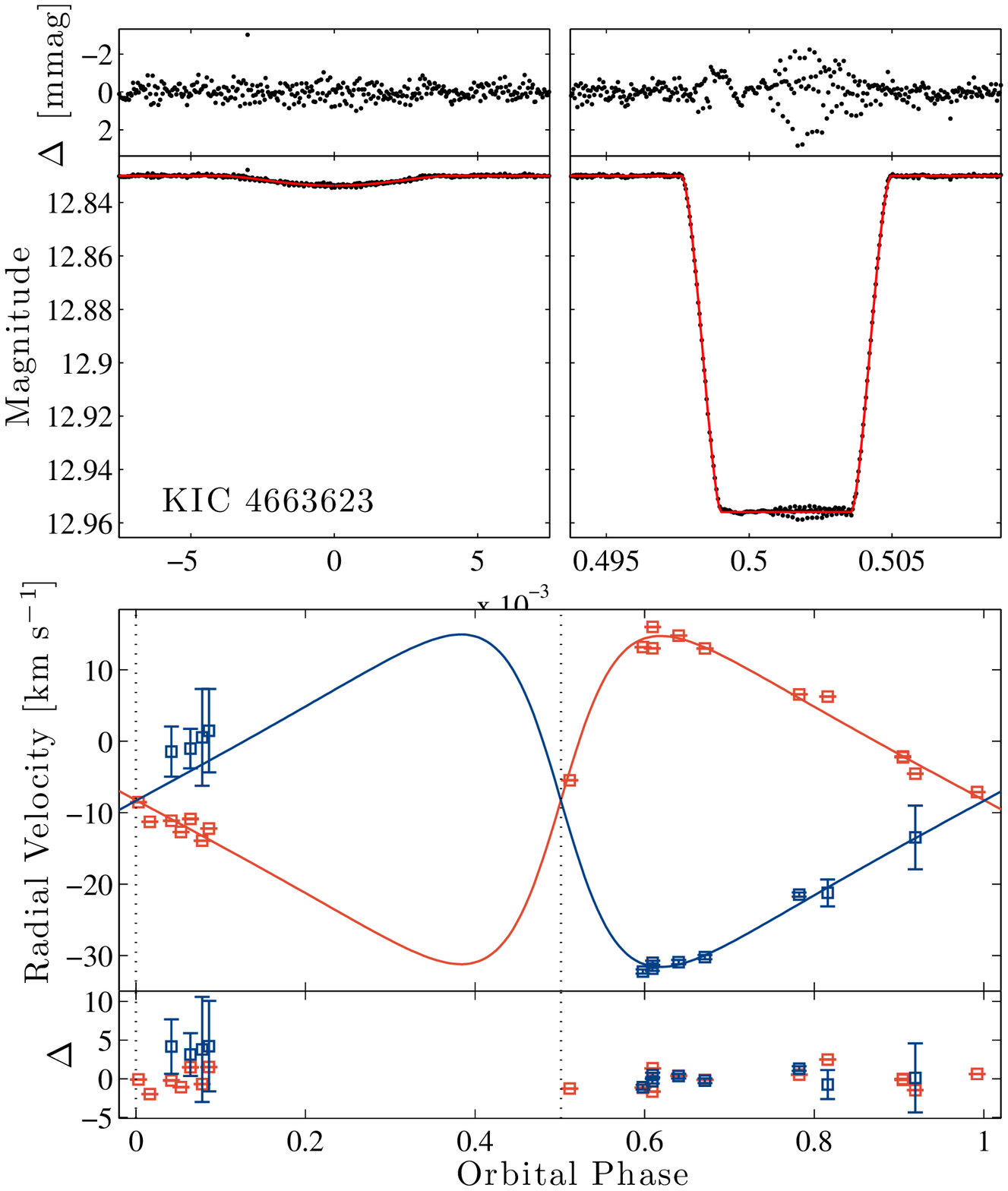}
\plotone{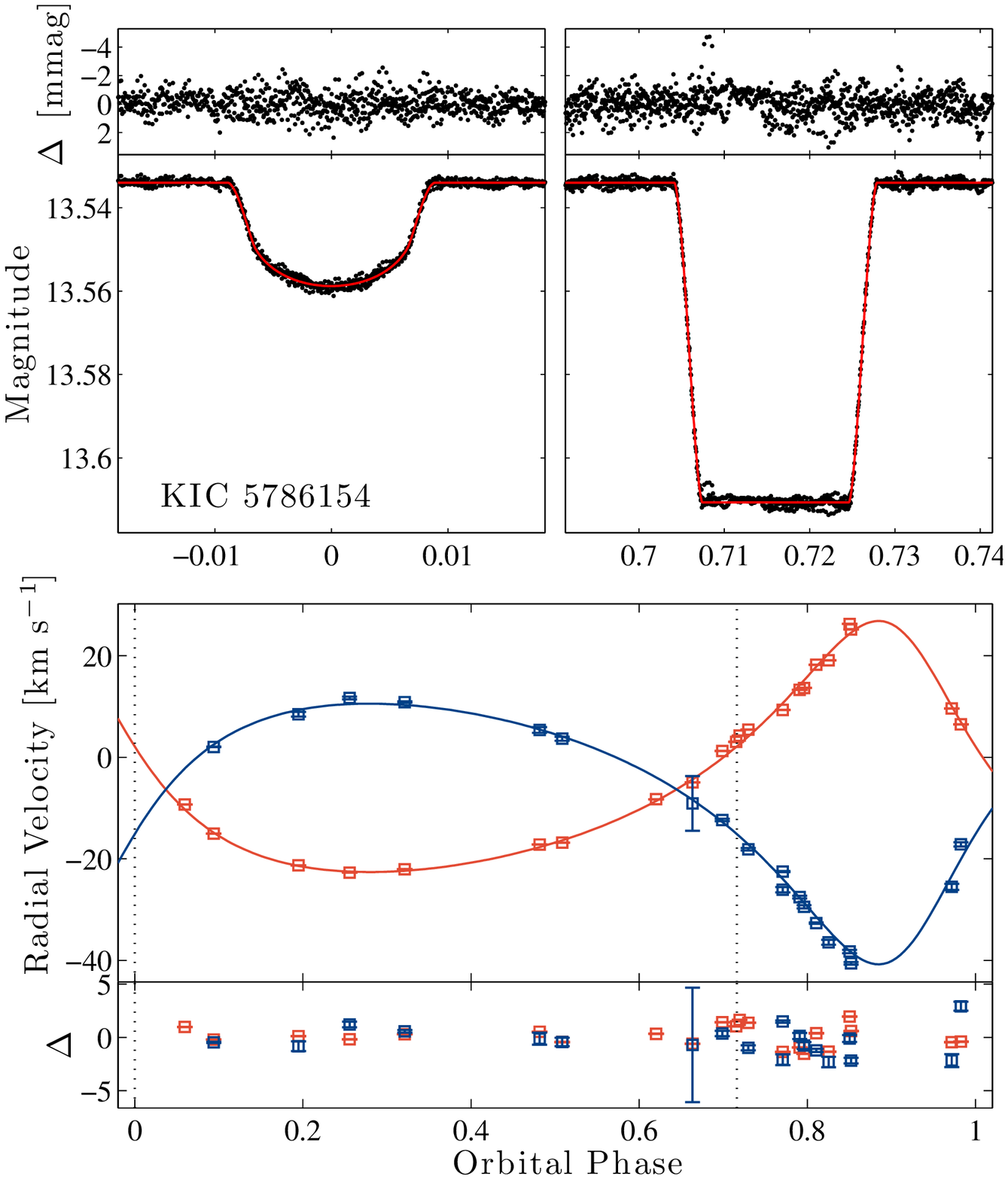}
\plotone{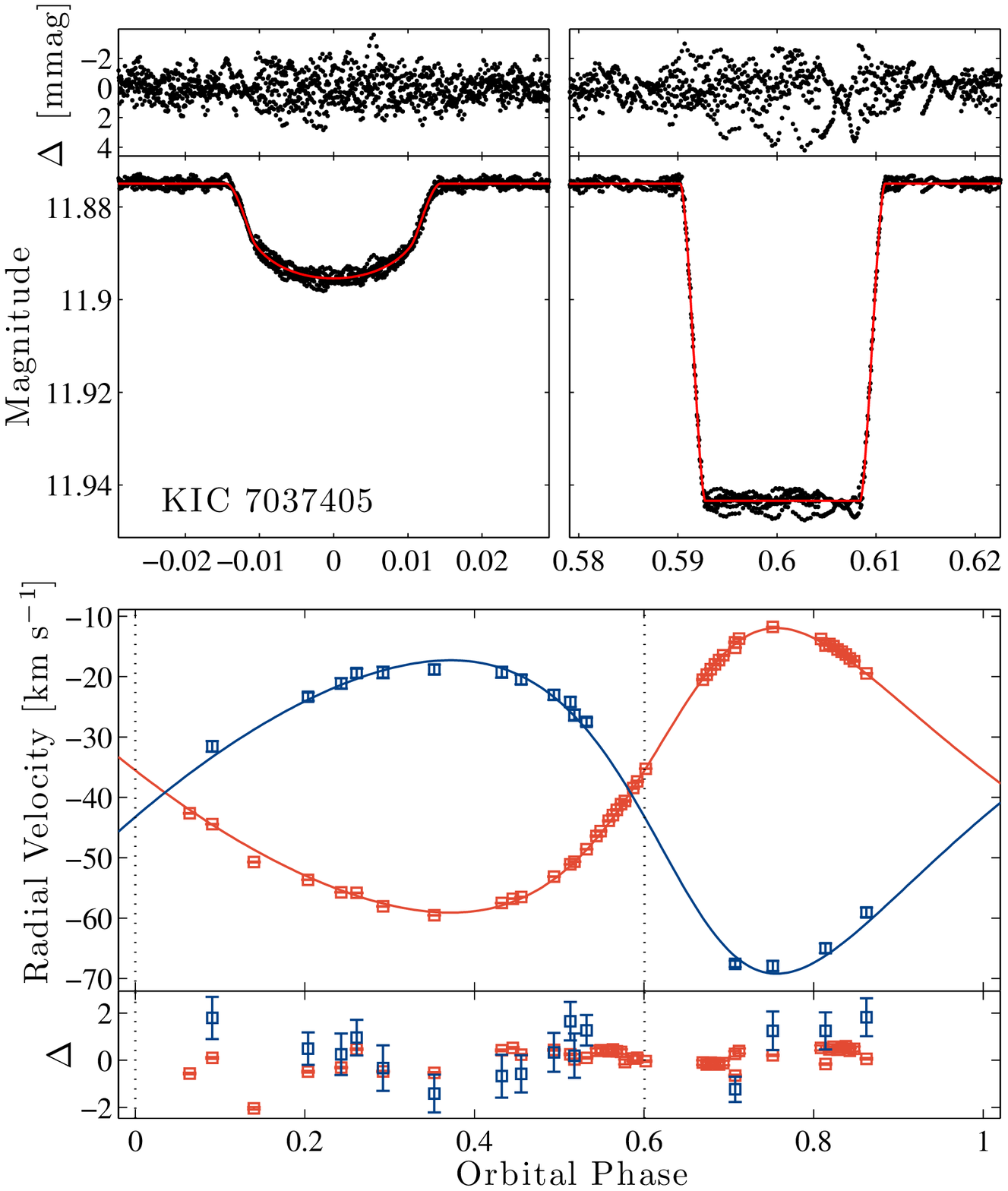}
\plotone{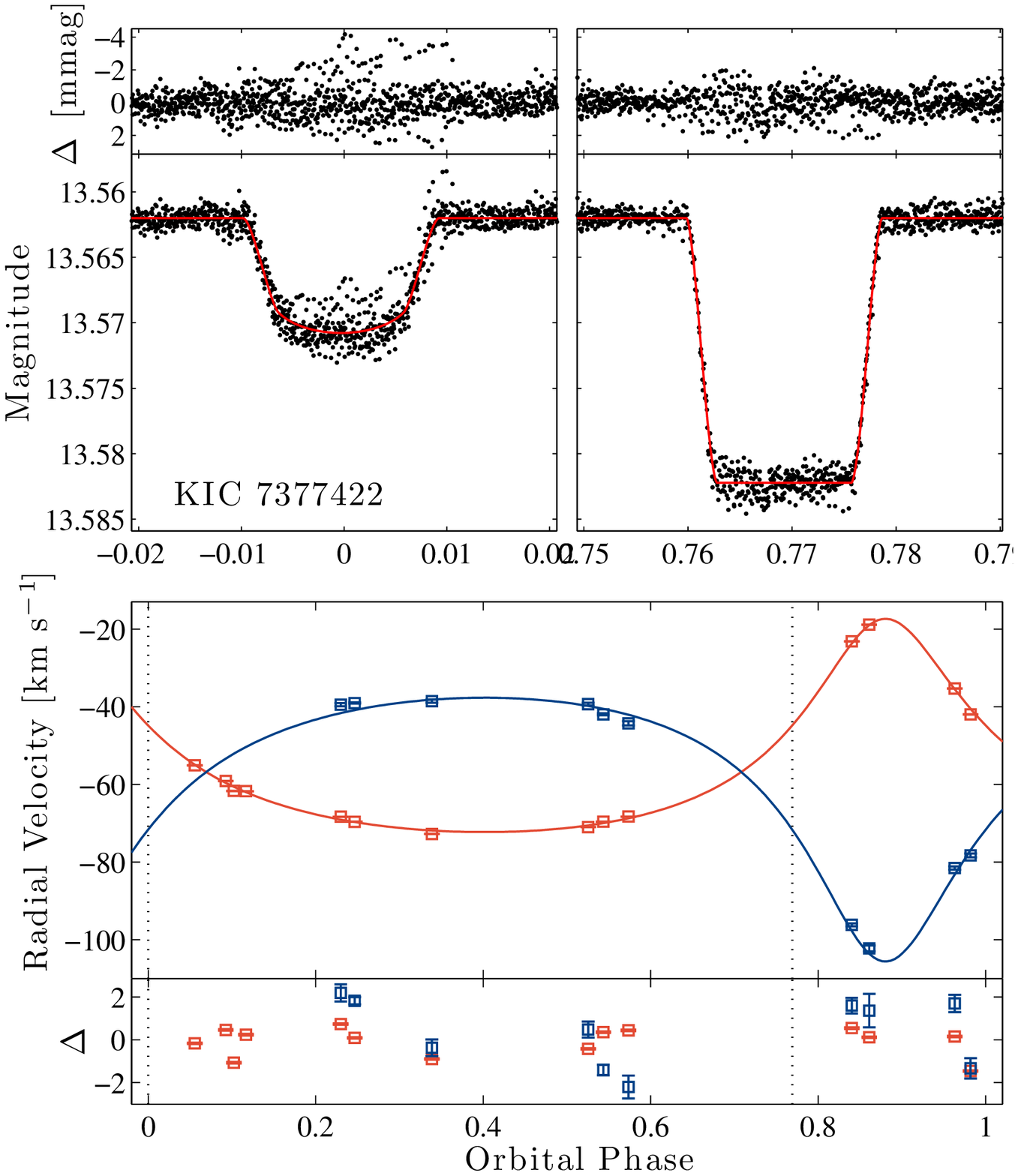}
\plotone{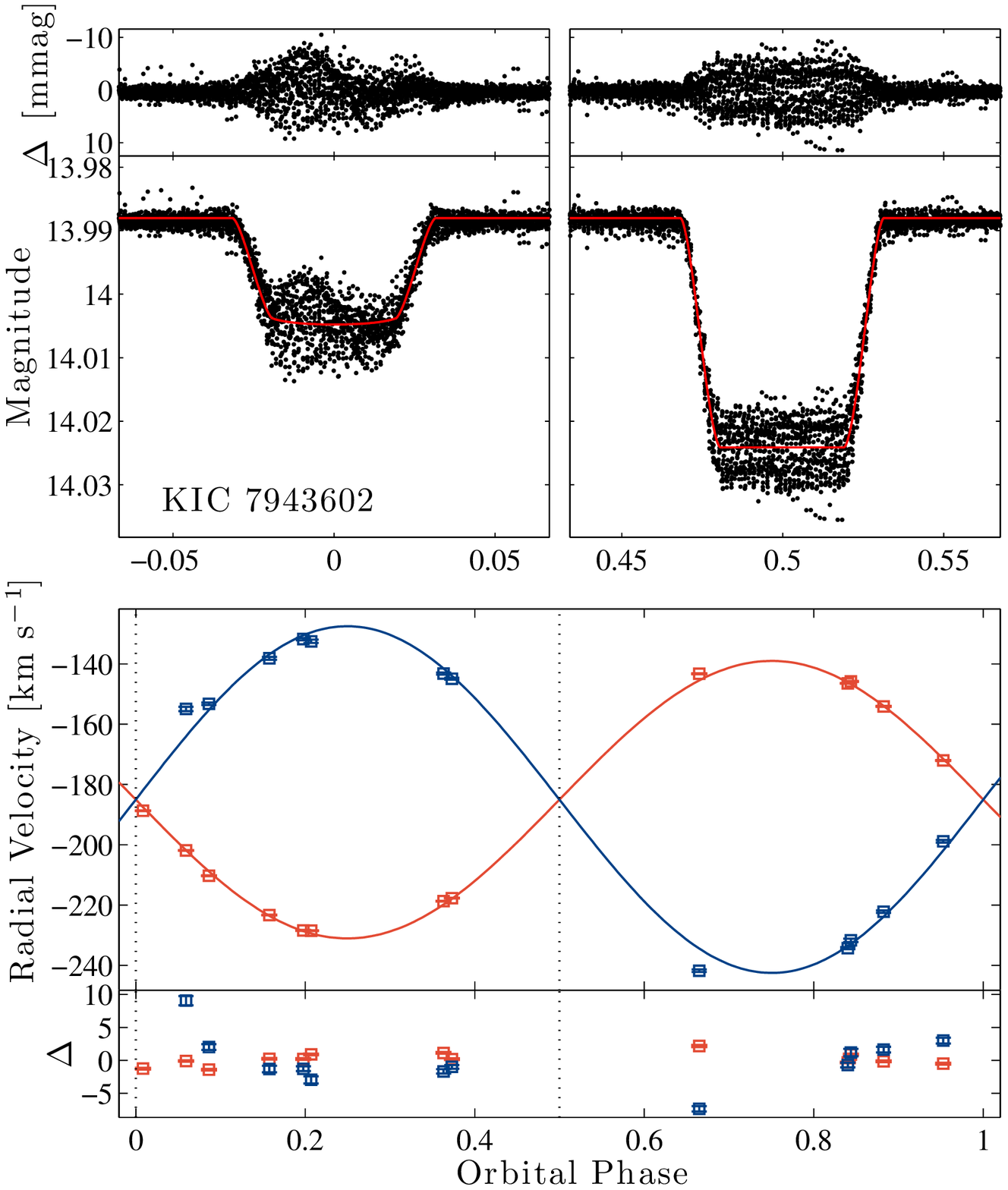}
\plotone{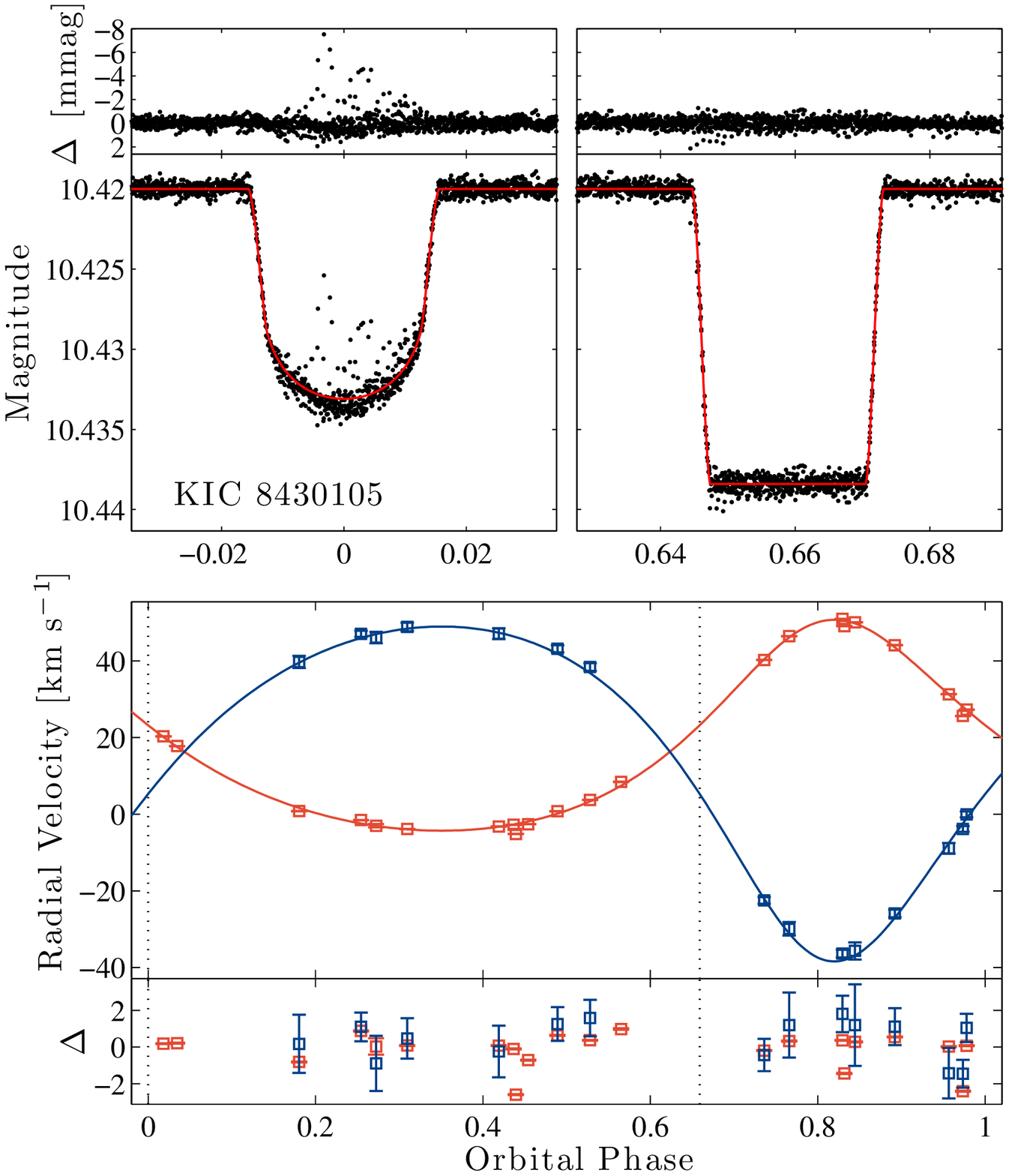}
\plotone{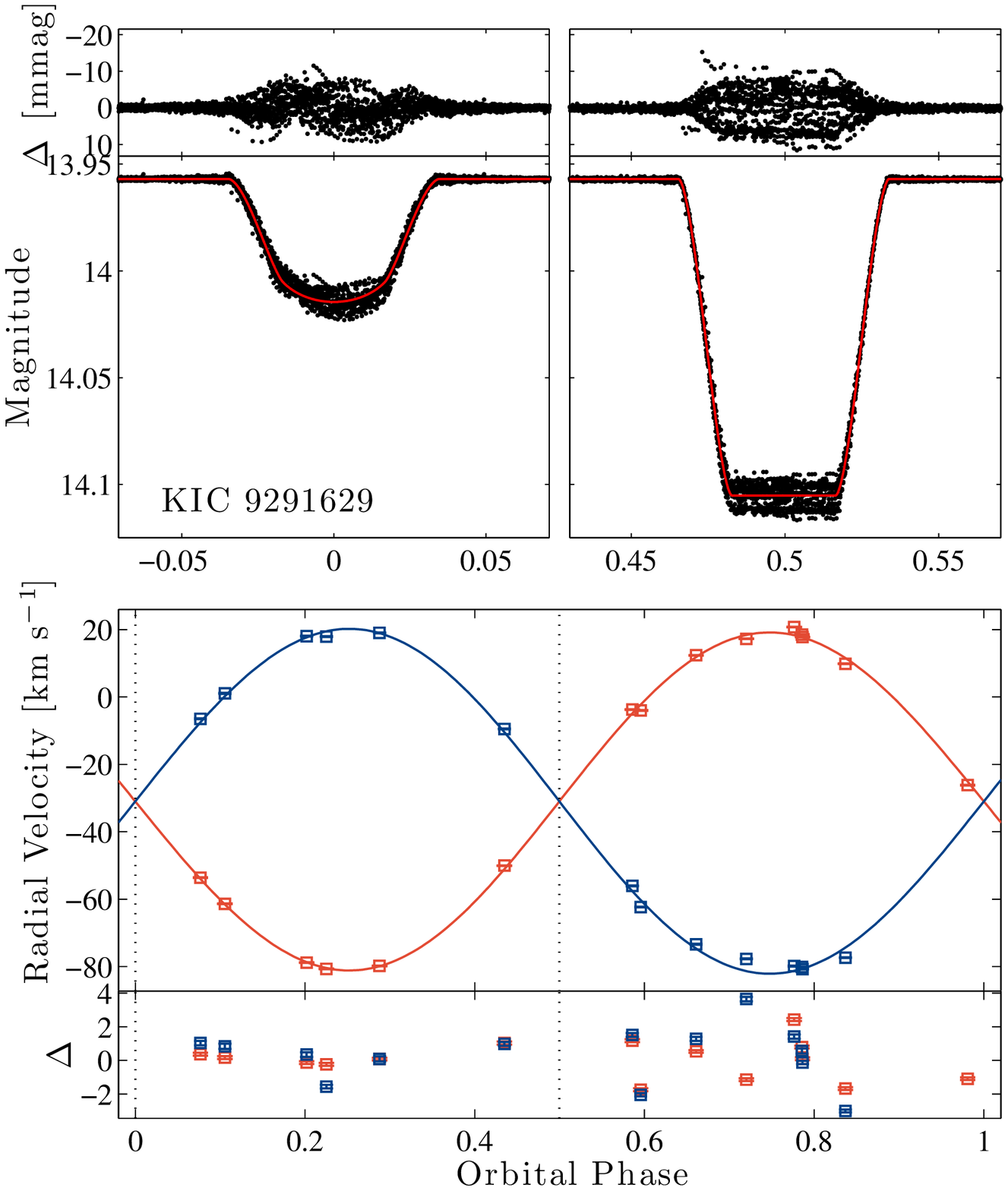}
\plotone{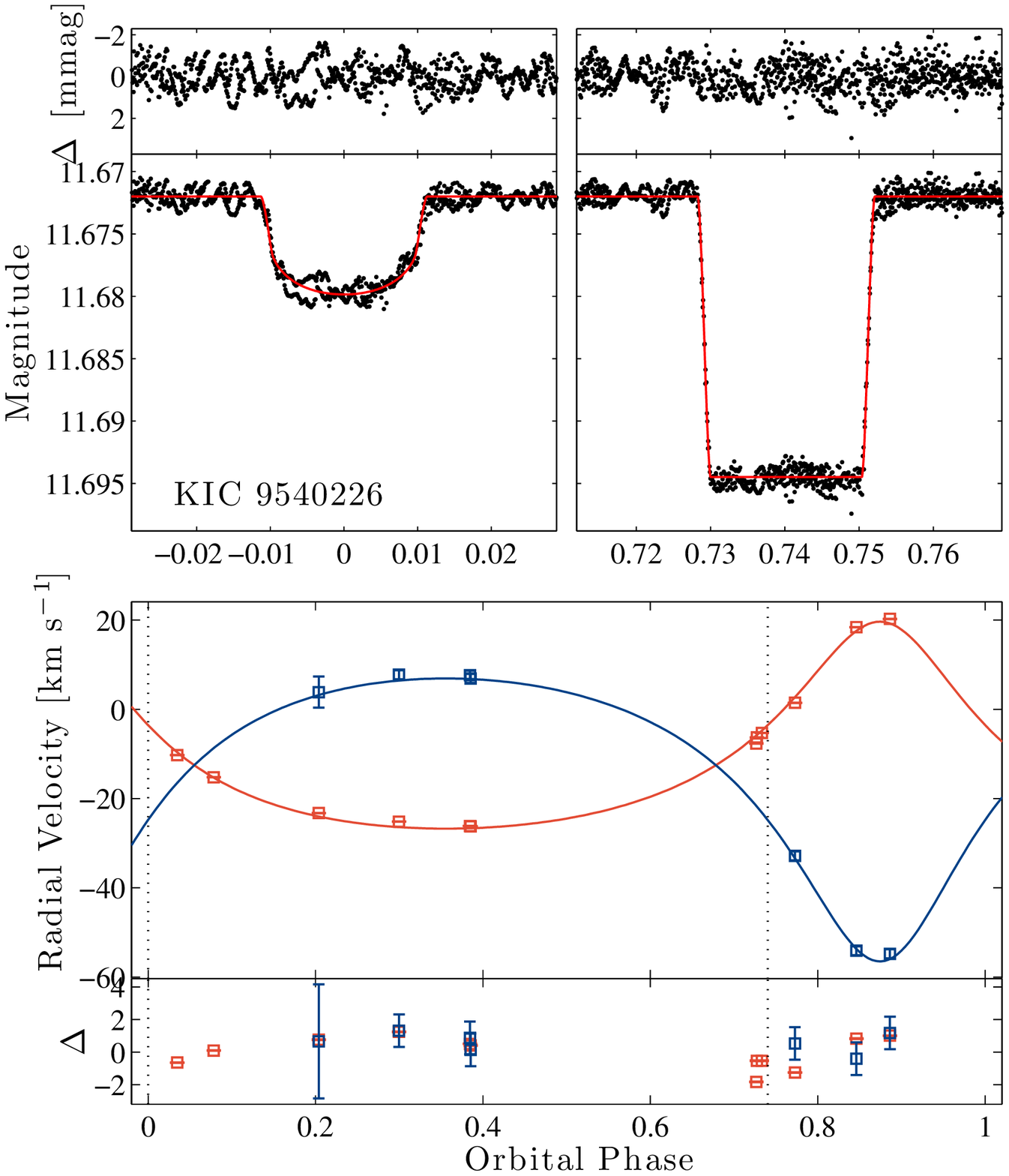}
\plotone{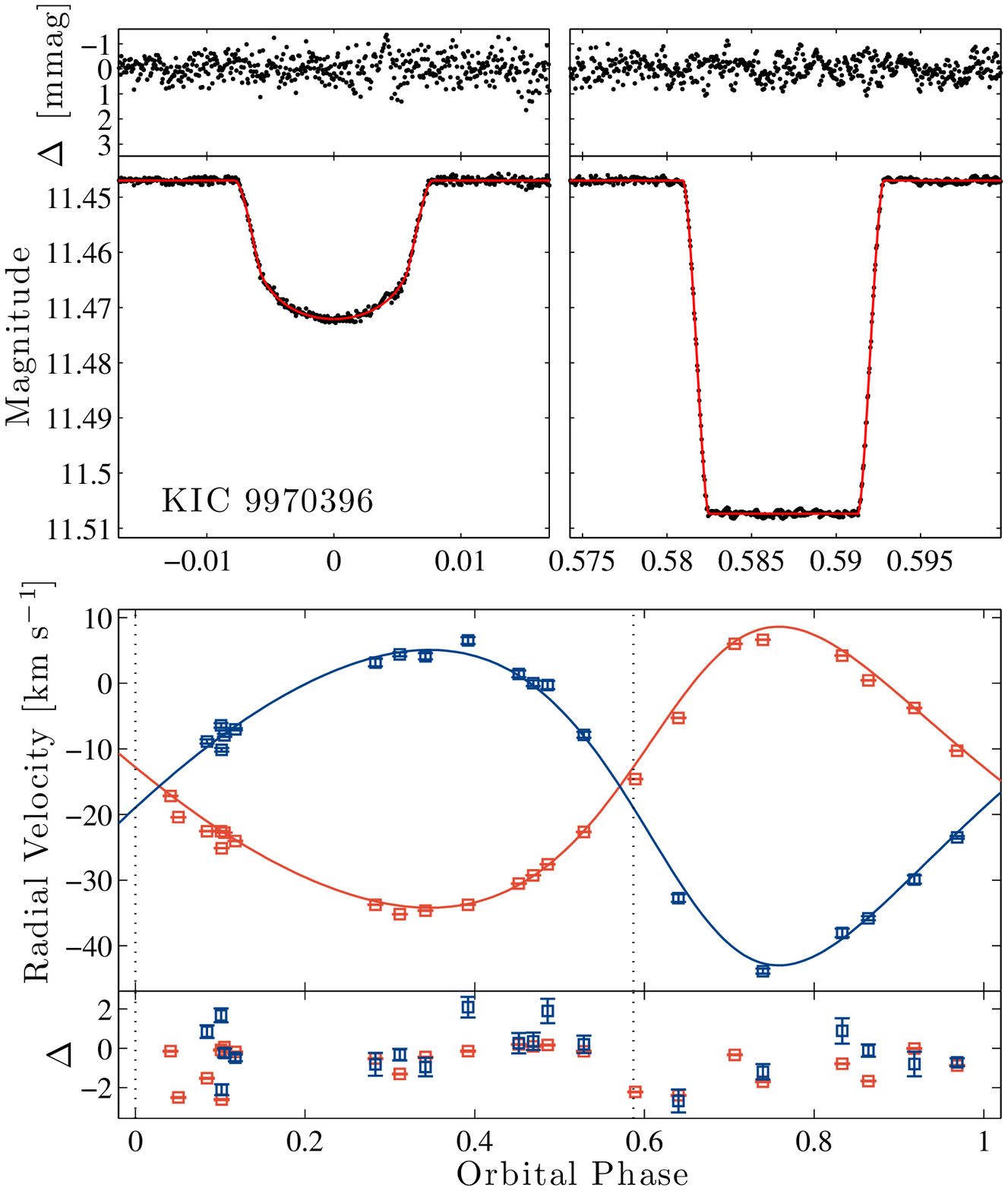}
\plotone{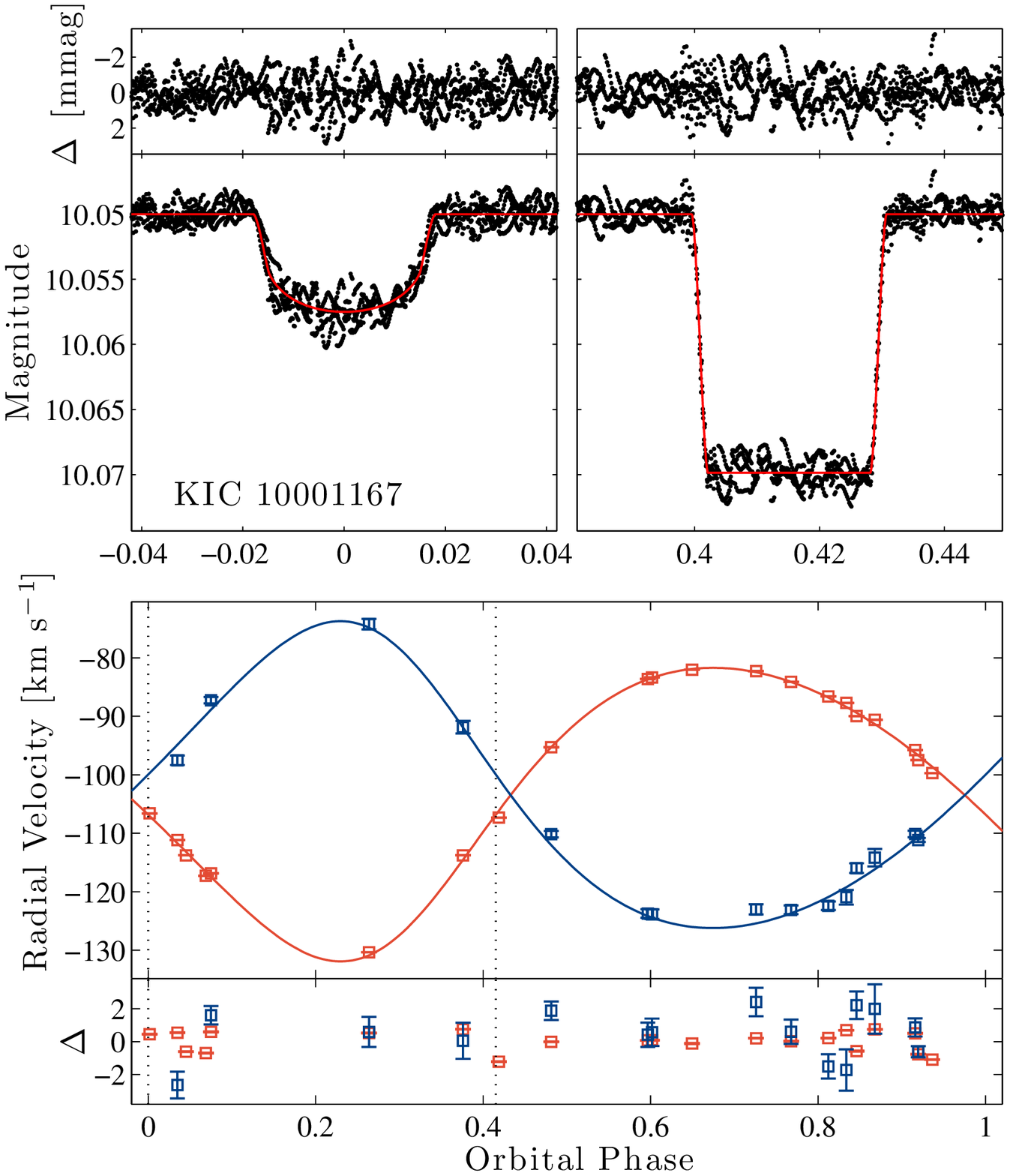}
  \caption{Combined modeling  of \textit{Kepler} light curves and RV measurements. All SB2 systems are represented here, except for KIC 9246715 which can be found in \citet{Rawls_2016}. \label{fig_ph_rv_sb2}}
\end{figure*}

\begin{figure*}[t]
\epsscale{0.33}
\plotone{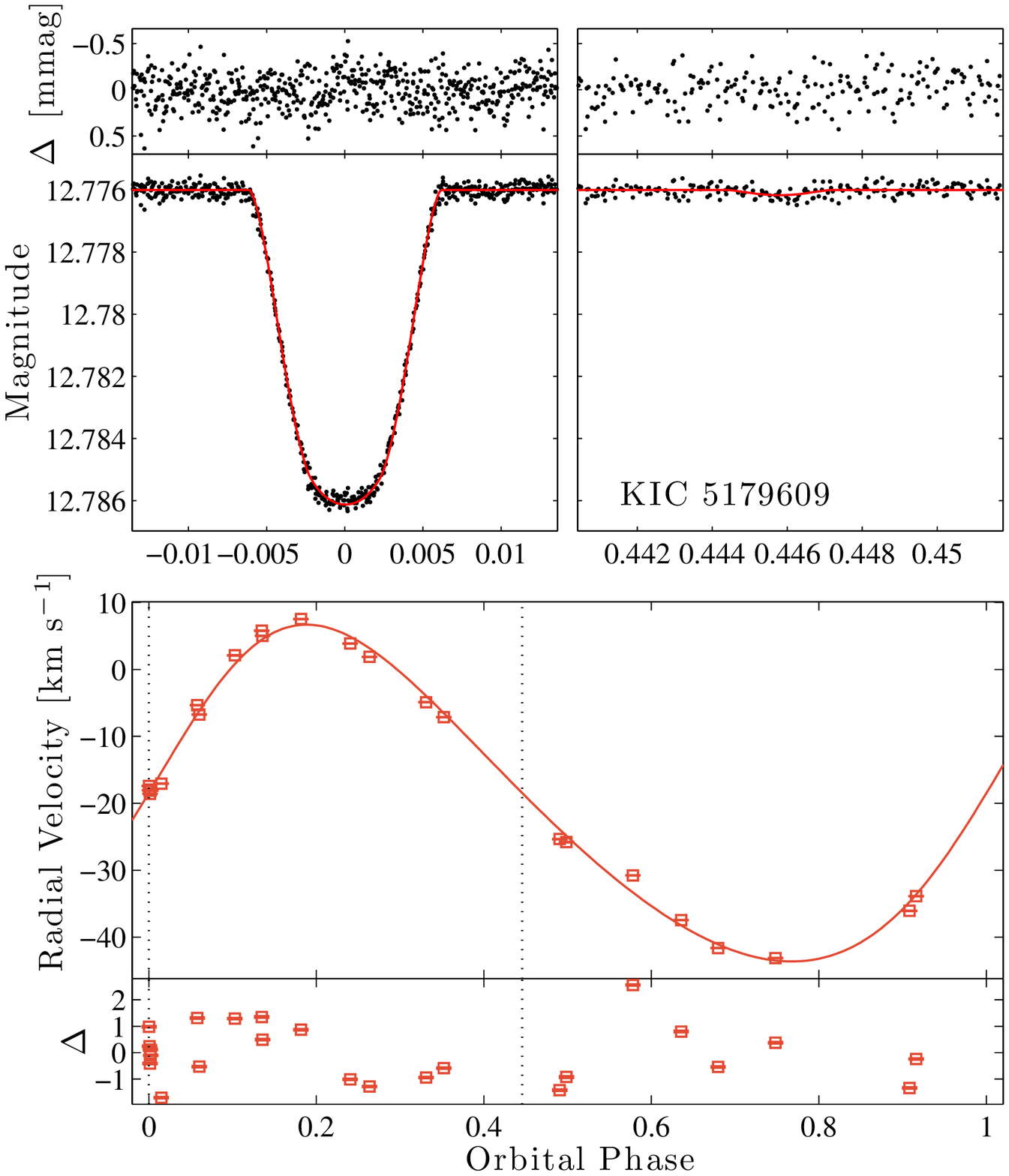}
\plotone{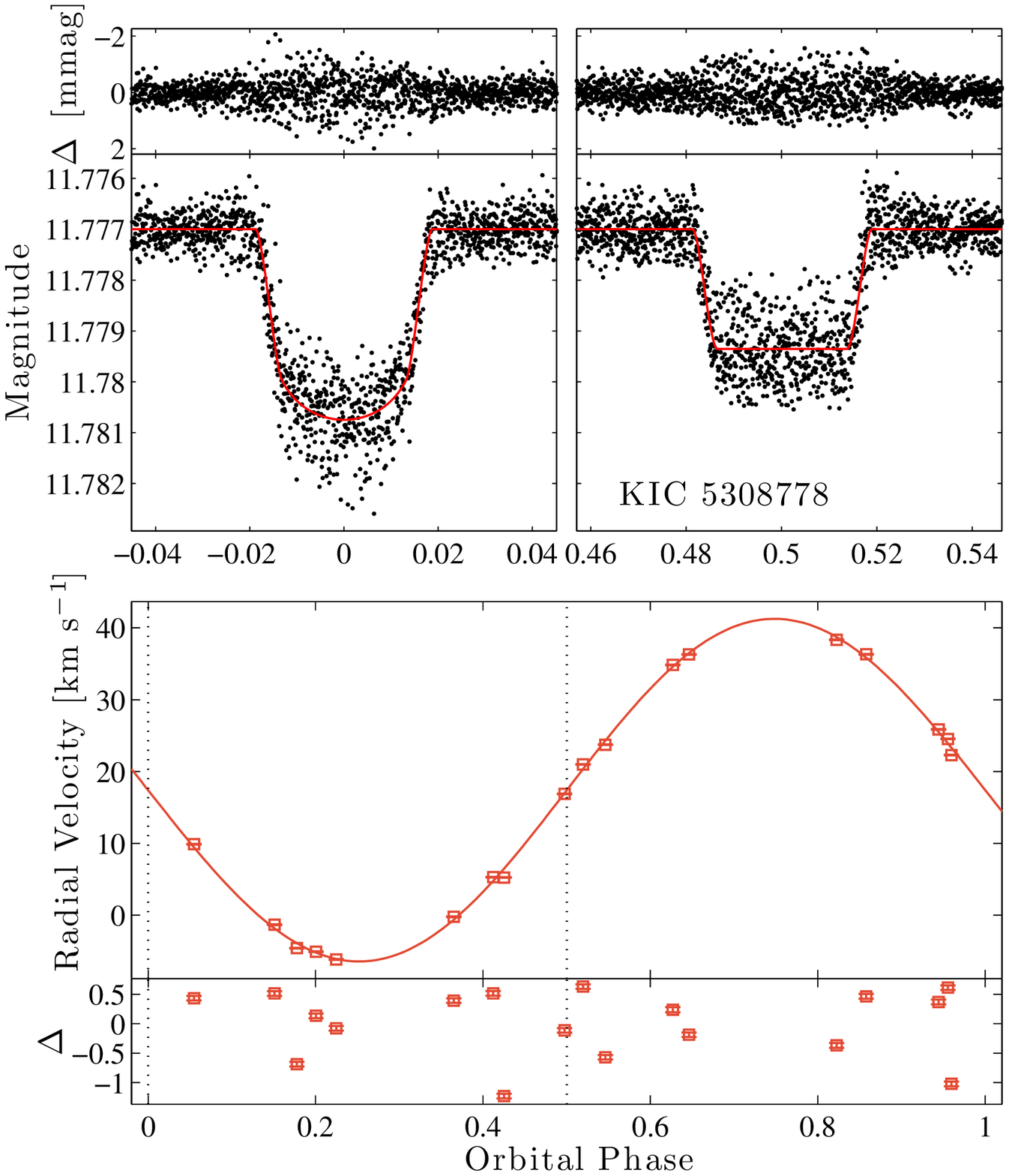}\\
\plotone{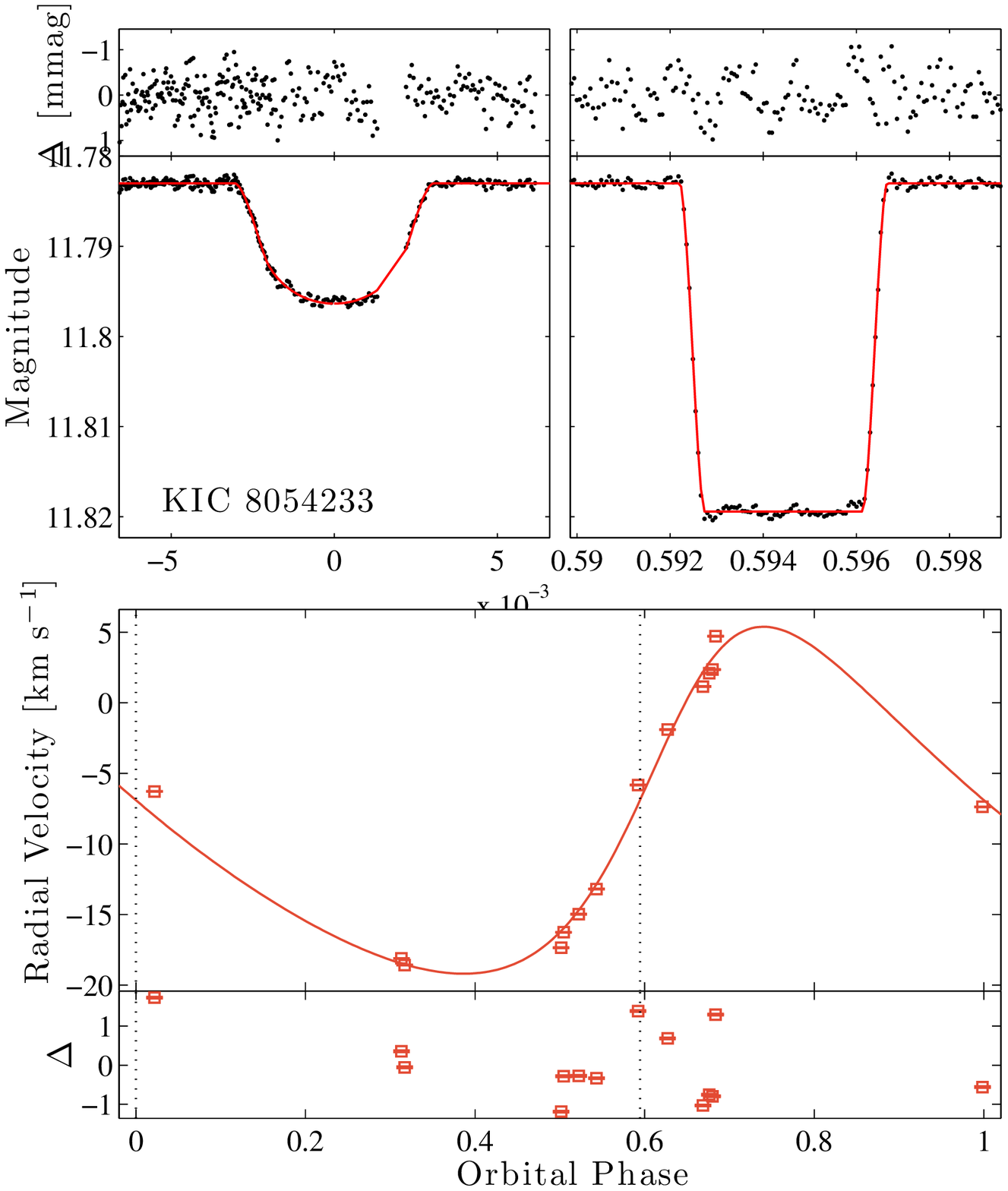}
\plotone{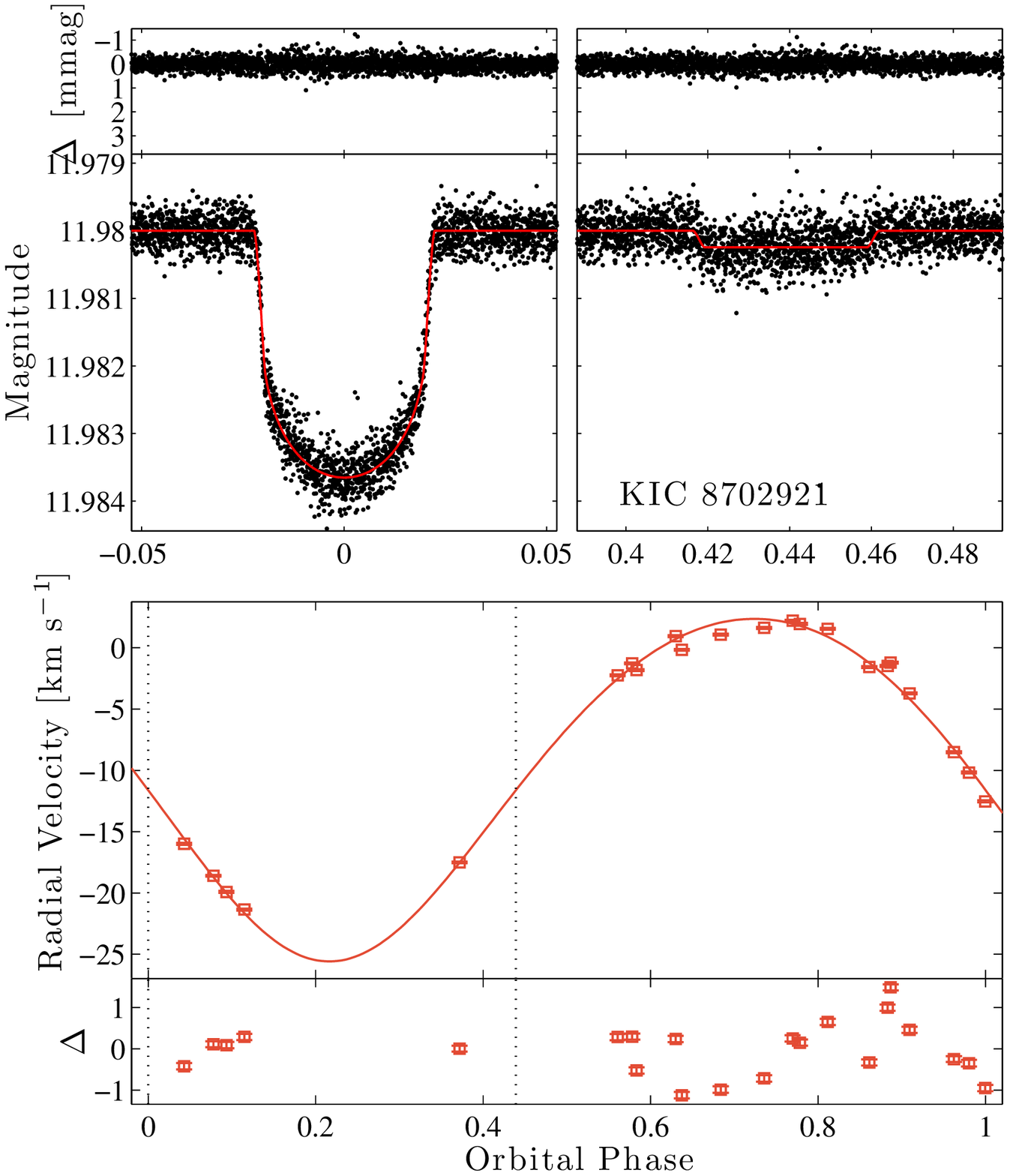}
  \caption{Combined modeling  of \textit{Kepler} light curves and RV measurements. All SB1 systems are represented here.\label{fig_ph_rv_sb1}}
\end{figure*}
%-----------------------------------------------------

Modeling the shape of the primary eclipse is highly dependent on the choice of the limb-darkening law of the stellar atmospheres.
Indeed, limb darkening is correlated with the eclipse's impact parameter (a function of inclination), ratio of radii, and surface-brightness ratio. In our specific sample, companion stars are always much smaller than the RG ($R\ind{cmp}/R\ind{RG} < 0.23$), so that the limb darkening law of the companion has a negligible influence on the eclipse shapes. In fact, many fits fail because the limb darkening parameters of the companion converge toward unphysical values ($<0$ or $>1$ at the limb). Therefore, we choose not to fit the limb darkening for the companion.  Instead we get a first estimate of the limb darkening (quadratic law) with the JKTLD routine, and then subsequently only fit the first-order term for the RG. Among various options, JKTLD is able to provide a limb-darkening estimates corresponding to the \textit{Kepler} bandpass by interpolating \citet{Sing_2010} coefficients.

%We estimate the error bars on the fitted parameters with the Monte Carlo option in JKTEBOP. The best-fitting light curve model is re-evaluated at the phases of the actual observations. Gaussian simulated observational noise is added and the result is refitted. This process is repeated and the range in parameter values found gives the uncertainty in that parameter.  For each of our systems, 1000 iterations were performed. Almost all models converged smoothly as all systems display total eclipses, except for KIC 5179609, where estimating its radius and temperature ratio was difficult. Data and models are plotted in Fig.~\ref{fig_ph_rv_sb2} for SB2 systems and and Fig.~\ref{fig_ph_rv_sb1} for SB1 systems.

As regards errors and noise, it is clear from the big scatter during eclipse in Figs.\ \ref{fig_ph_rv_sb2} \& \ref{fig_ph_rv_sb1} that the dominant noise is systematic (due to the pulsations and/or spots) and not Poisson. The noise is thus correlated from point to point, on timescales of several hours. We therefore estimate the error bars on the fitted parameters with Task 9 in JKTEBOP, which takes as input a parameter file, finds the best fit, and then assesses the errorbars on the parameters of the fit in a way which accounts for correlated noise. The residual-shift method is used, where the residuals around the best fit are shifted point-by-point through the observational data. After each shift a new best fit is calculated. Approximately 1000 iterations were performed for each fit, which corresponds to shifts much longer than the oscillation or spot timescales. The 1-$\sigma$ errors are calculated by sorting the best fits and taking values which correspond to the central 68.3\,\%. All models converged smoothly as all systems display total eclipses, except for KIC 5179609, where estimating its radius and temperature ratio was difficult.

For SB2s, the masses of both components are directly determined from the combined modeling of RV and photometric measurements. Note that masses can be estimated with good precision from RVs only for EBs with relatively distant stars ($\sin i \approx 1$). The mass of the RG is determined by the motion of the companion star, and vice versa. Since error bars on masses are related to the dispersion of RV measurements and the RG spectral lines display a much larger SNR, the precision on masses is better for companion stars than RGs.  We obtain a median precision on RG masses of $3.35\,\%$ (from 1.4\,\% for 9970396 to 19.1\,\% for 4663623), and of $2.55\,\%$ for companions (from $0.7\,\%$ to $11.4\,\%$).

The relative precision on the radii is better than for the masses because of the exquisite quality of the \textit{Kepler} light curves.
There is no difference of precision for RGs and companions, and we report a median precision of 1.04\,\% (from 0.4 to 5.1\,\%). However, the accuracy of the radius measurements is lower than for masses, as it depends on stellar atmospheric models, i.e., limb darkening, which is not necessarily well constrained and can be the cause of small biases. Also, contamination from a third star in the aperture can affect the radius estimates through the ratio of the radii.

For most systems, we have fewer measurements of the companion's Doppler shift since it is difficult to track its lines in the vicinity of eclipses. The median number of RV measurements for RG lines is 15.5 (from 12 to 44) and 13 for companions (from 7 to 19). The phase coverage is a function of the weather at APO and also of orbital periods. In addition, the two systems with the largest orbits (the SB1 8054233 and SB2 4663623) were discovered about two years after the rest of the sample, resulting in poorer phase coverage and precision.

The four SB1 systems are worthy of discussion, despite not being suitable for testing asteroseismology. By assuming asteroseismology provides reliable RG masses and radii, it is possible to estimate the companion masses and radii. If the Doppler shift is measured only for the RG, the relation
\begin{equation}
\frac{\left(M_2 \sin i\right)^3}{\left(M_1 + M_2\right)^2}\ =\ \frac{K_1^3 P \left(1-e^2\right)^{3/2}}{2 \pi G}
\label{mass_func}
\end{equation}
allows us to retrieve the companion's mass, where $M$ are stellar masses, the subscript $1$ refers to the star for which we measure RVs, $K_1$ is the amplitude of the RV, and G the gravitational constant. The mass ratio $q = M_2/M_1$ is one of the three roots of the equation:
\begin{equation}
q^3 - \alpha q^2 - 2\alpha q - a = 0,
\end{equation}
where
\begin{equation}
\alpha = \frac{1} {2 \pi G} \frac{K_1^3 P \left(1-e^2\right)^{3/2}}{M_1 \sin^3 i}.
\end{equation}
Note this method can be used to determine the mass $M_2$ of a planet transiting its host star, whose mass $M_1$ is estimated independently.

%===============================================================================================
\subsection{Physical properties from asteroseismic scaling relations and mixed modes}
\begin{figure}[t]
\epsscale{1.15}
\plotone{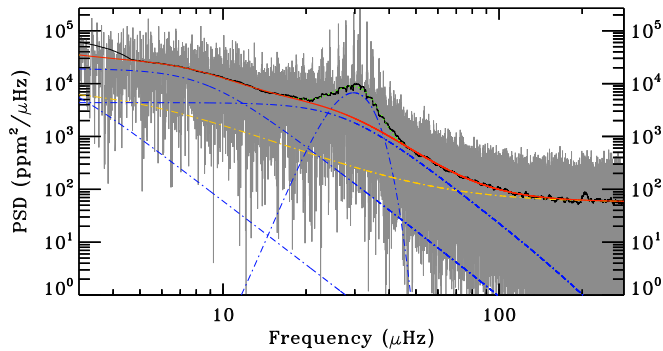}
\caption{Background fit of KIC 5786154, as derived by \textsc{D\large{iamonds}}. The original PSD is shown in gray, while a smoothed version with boxcar width set to $5\,\Delta\nu$, is shown as a black line to guide the eye. The red thick line represents the background model without the Gaussian envelope, while the green dotted
line accounts for the additional Gaussian component. The individual components of granulation and long-trend variation are shown by blue dot-dashed lines. The yellow dot-dashed line shows the superposition of white and colored noise. \label{fig_bkgd}}
\end{figure}

The stellar variability (starspots, granulation, etc.) is the cause of what is usually called the stellar ``background'' in the frequency domain. To determine the global asteroseismic parameters, we fit the background with a sum of two super Lorentzian functions centered on zero frequency, a Gaussian accounting for the mode envelope, and white noise. The center of the Gaussian function constitutes our measurement of $\nu\ind{max}$. Given the importance of $\nu\ind{max}$, both co-authors Gaulme and Corsaro independently fitted this parameter and found values that are fully compatible. Gaulme used a routine based on a Bayesian maximum a posteriori method \citep{Gaulme_2009}, while Corsaro used the Bayesian DIAMONDS pipeline \citep{Corsaro_2014} following the methodology explained by \citet{Corsaro_2015} and illustrated in Fig.~\ref{fig_bkgd}. The large frequency spacing $\Delta\nu\ind{obs}$ is obtained in two steps: we get a first estimate by measuring the maximum of the envelope of the autocorrelation of the time series filtered in the frequency range corresponding to the oscillations \citep{Mosser_Appourchaux_2009}. However, the presence of many mixed $\ell=1$ modes in red giants, which are not equally spaced in frequency, may bias autocorrelation methods. As a second step, we compared this first estimate with the so-called universal RG pattern, which should rectify any potential bias caused by $\ell=1$ modes \citep{Mosser_2011}. We are aware that measuring $\Delta\nu\ind{obs}$ by comparing the $\ell=0$ frequencies to those predicted by the universal pattern, erases any diversity that may exist in the RG oscillation spectra. Given the objective of the present paper is to carefully compare masses and radii, we chose to re-estimate all the global asteroseismic parameters, even though it was done by \citet{Gaulme_2014}. The main difference consists of two independent estimates of $\nu\ind{max}$ and the use of the universal RG pattern for $\Delta\nu$.

%------------------------------------
%------------------------------------

The $\Delta\nu$ scaling relation comes from the fact that the oscillation pattern of low-degree pressure modes can be described by a second-order relation \citep{Tassoul_1980}. The first-order term shows that modes of a given degree are evenly spaced in frequency as a function of their radial order $n$. The second-order term is responsible for the curvature observed in the \'echelle diagrams used for analyzing the oscillation spectra. This approximate relation that describes the mode frequencies is called asymptotic, since its derivation is strictly valid only for large $n$. The common use of asteroseismic scaling relations \citep[e.g.][]{Chaplin_2011c,White_2012} neglects the curvature and assumes the large frequency spacing that is observed, $\Delta\nu\ind{obs}$, to be equal to the asymptotic large spacing $\Delta\nu\ind{as}$, for which the scalings are correct \citep[but only in the case of homologous stars, e.g.,][]{Belkacem_2013}. In the case of RGs with $\nu\ind{max} \leq 50\ \mu$Hz, the oscillations' radial orders are less than ten, and the first-order asymptotic development is not assured to be valid.
\citet{Mosser_2013} proposed a semi-empirical relation to convert the observed into the asymptotic spacing: $\Delta\nu\ind{as} = \Delta\nu\ind{obs} (1 + \zeta)$, with $\zeta = 0.038$ and where the reference solar values are modified to $\Delta\nu_\sun=138.8\ \mu$Hz and $\nu\ind{max\odot}=3104\ \mu$Hz, instead of the observed values of 134.9 and 3050~$\mu$Hz. The use of the second-order development leads to smaller masses and radii by 8.4\,\% and 3.5\,\% for red-giant oscillators. 

The transformation of $\Delta\nu\ind{obs}$ into $\Delta\nu\ind{as}$ proposed by Mosser et al. is actually debated \citep{Hekker_2013}, whereas it is generally admitted that asteroseismic scaling relation overestimate masses. Other studies have introduced empirically-calibrated corrections to the scaling laws \citep[e.g.,][]{White_2011, Miglio_2012, Sharma_2016, Guggenberger_2016}. \citet{White_2011} applies a correction on $\Delta\nu$ based on numerical models, which is a function of the effective temperature and increases it by less than 1\,\% for a red giant, and makes use of \citet{Broomhall_2009}'s $\nu\ind{max,\odot}=3100~\mu$Hz. \citet{Sharma_2016} proposed a correction on $\Delta\nu$ for red giants based on stellar models, which is a function of the evolutionary stage (RGB, RC), $\nu\ind{max}$, $\Delta\nu$, $T\ind{eff}$, and [Fe/H]. \citet{Guggenberger_2016} also applies a model-based correction to the reference solar $\Delta\nu_\odot$ that is a function of metallicity and temperature. For most of the red giants in our sample, the modified  reference $\Delta\nu_\odot$ using their correction yields values less than the typical ${\rm 135.1\,\mu Hz}$. Their models also assume $\nu\ind{max,\odot}=3050~\mu$Hz. These three attempts tend to decrease masses and radii with respect to the original scaling relations. The \citet{Miglio_2012} approach was applicable for two specific open clusters and aimed to quantify the effects of mass loss on the RGB to the horizontal branch. It is not suitable for our targets. Alternatively, some users have slightly increased the reference $\nu\ind{max,\odot}$, to ensure that $\nu\ind{max}$ has been consistently measured with the same method for both the Sun and the \textit{Kepler} stars, and this makes masses and radii decrease too. For example, \citet{Kallinger_2010} and \citet{Chaplin_2011c} have introduced a $\nu\ind{max,\odot}$ equal to 3120 and 3150 $\mu$Hz, respectively, which has a significant influence on mass given its cubic dependence on $\nu\ind{max}$.  The masses drop 6.6 and 9.2\,\% respectively, while the radii drop  2.3 and 3.2\,\% with respect to the \citet{Kjeldsen_Bedding_1995} reference. In our tables, we report the asteroseismic values using \citet{Mosser_2013}'s corrections, but we also comment on the difference with the other scaling relations (see Fig.~\ref{fig_scalings}).

%-----------------------------------------------------
\begin{figure*}[t]
\epsscale{0.35}
\plotone{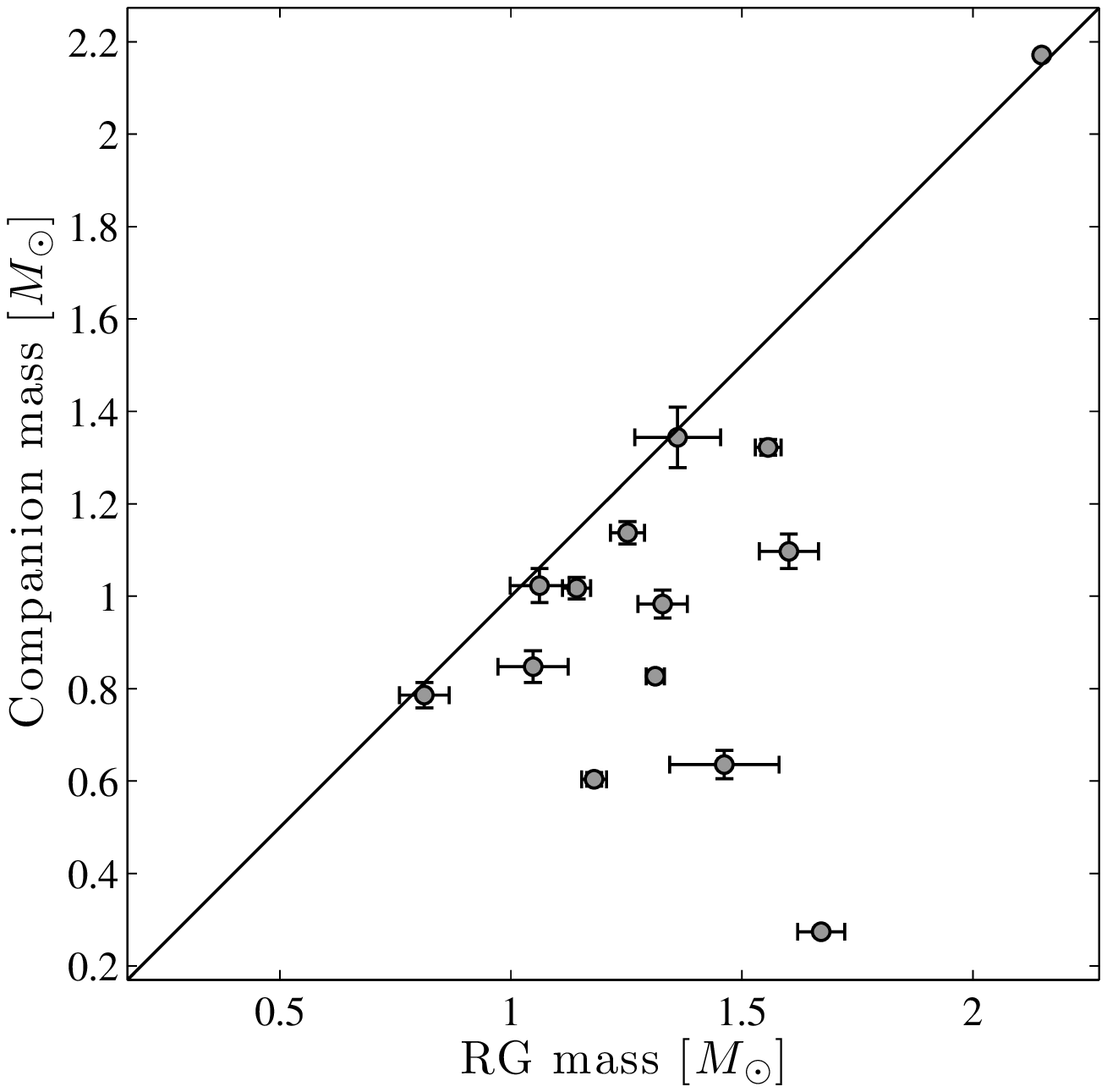}
\plotone{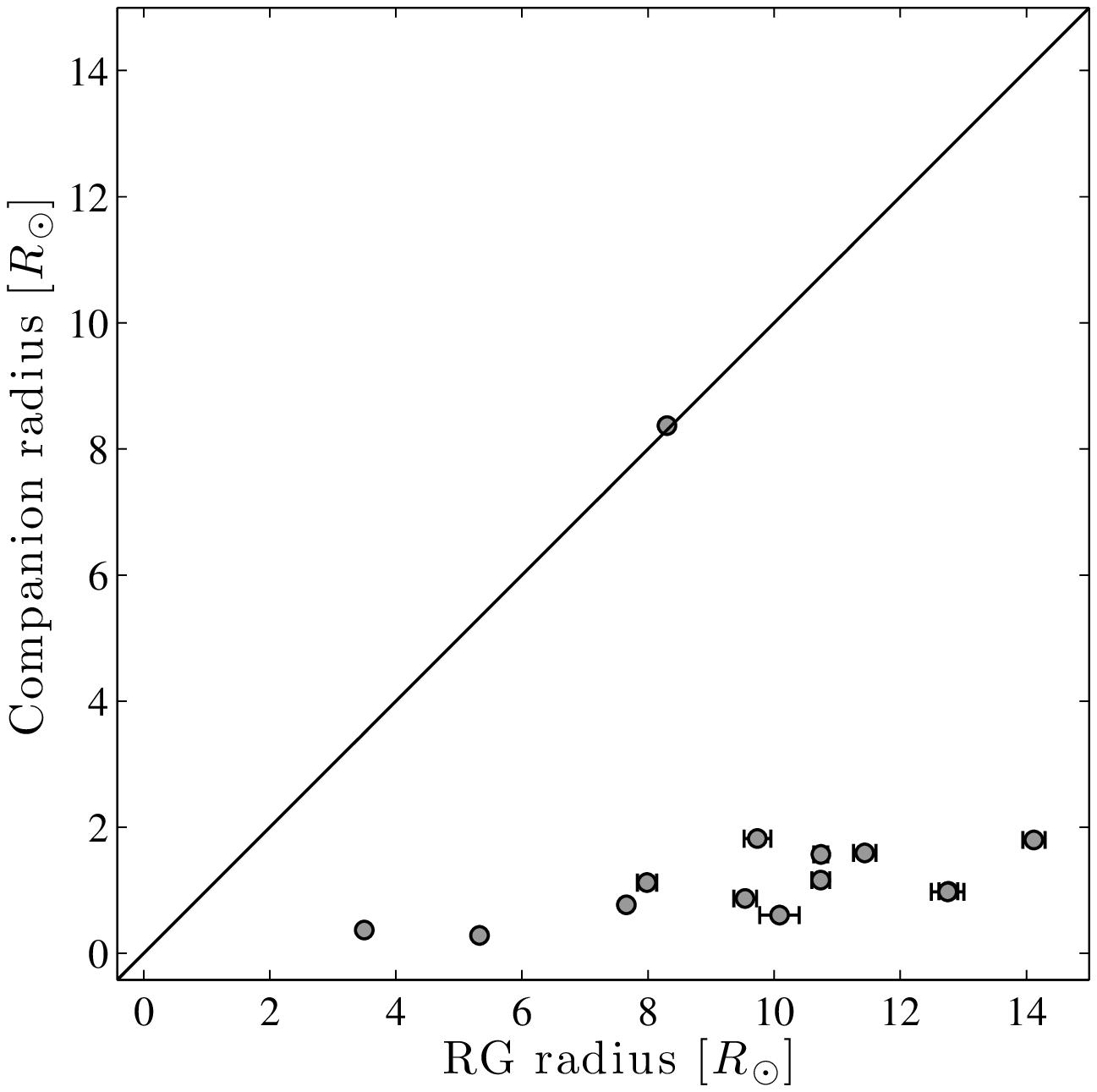}
\plotone{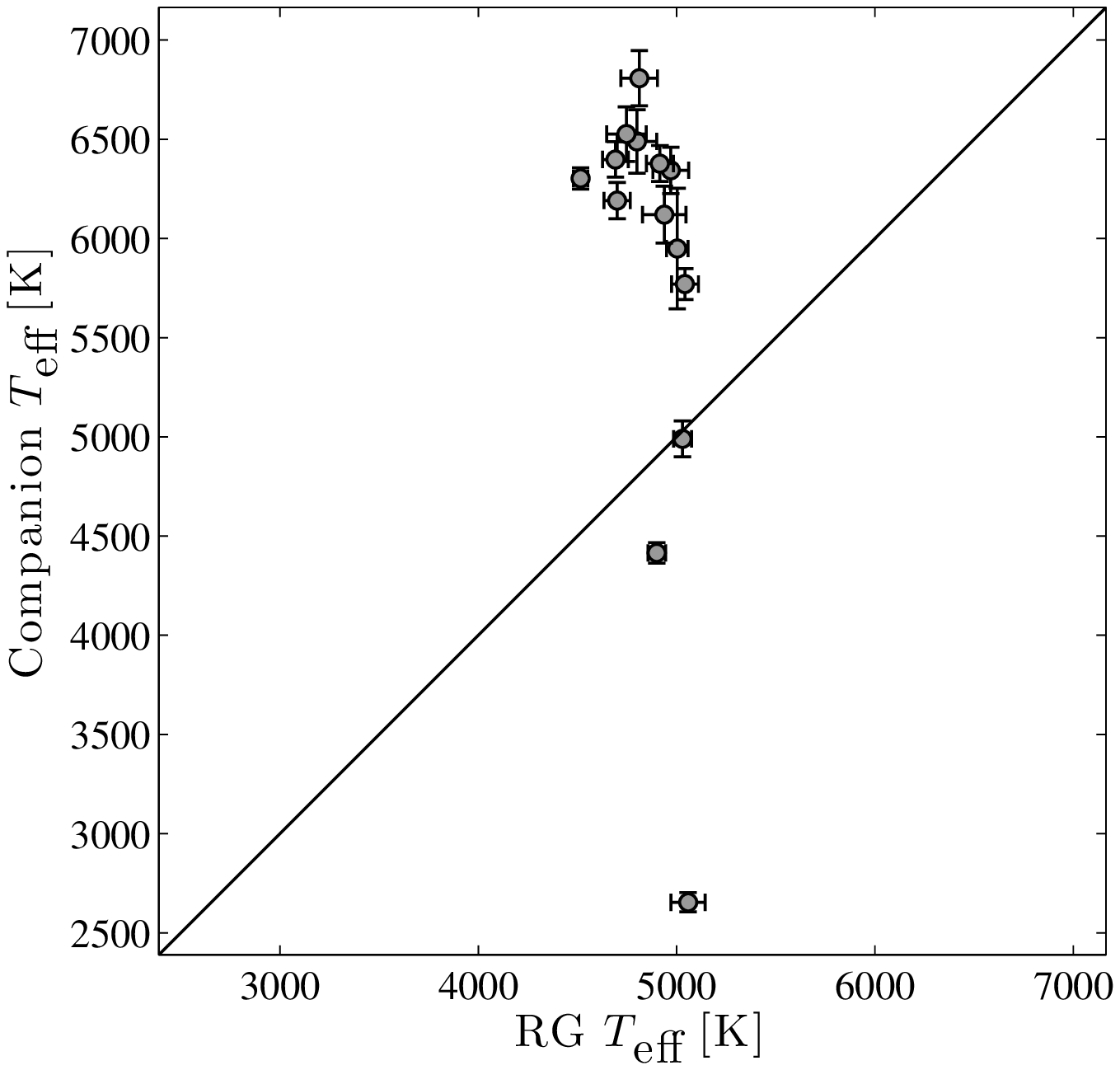}
\caption{Red giant versus companion  physical parameters. For SB1s, the masses and radii of the companion stars were obtained by combining asteroseismic masses and radii with the dynamical mass function (Eq.~\ref{mass_func}). \label{fig_RG_cmp}}
\end{figure*}
%-----------------------------------------------------

The evolutionary stage of RGs -- red-giant branch (RGB), main red clump (RC), secondary red clump (RC2), asymptotic giant branch (AGB) -- may be deduced from the study of mixed modes measured in the power spectra \citep{Bedding_2011}. When mixed modes are not detected, it is still possible to infer the nature of an RG based on $\Delta\nu$, or on mass and radii criteria \citep{Mosser_2012a}, or on $\Delta\nu$ and $\epsilon$ criteria \citep{Kallinger_2012}. Moreover, as suggested by \citet{Beck_2014}, an RC or AGB identification can be discarded for a star with mass lower than $1.8\ M_\odot$ if the separation between companions is less than $200\ R_\odot$ along its orbit, because it corresponds to the size that a low-mass star reaches at the tip of the RGB. The engulfment of the RG's companion would occur before reaching the RC stage. This analysis was performed by \citet{Gaulme_2014} for all the stars of this sample. In summary, KICs 9246715 \citep{Rawls_2016} and 8054233 are the only definite clump stars. KIC 8410637, which we include in our analysis, is possibly a clump star too, but its nature is unclear \citep{Hekker_2010,Gaulme_2014,Brogaard_2016}. All of the others are RGB stars.

%%%%%%%%%%%%%%%%%%%%%%%%%%%%%%%%%%%%%%%%%%%%%%%%%%%%%%%%%%%%%%%%%
%                                                                                    III. RESULTS
%%%%%%%%%%%%%%%%%%%%%%%%%%%%%%%%%%%%%%%%%%%%%%%%%%%%%%%%%%%%%%%%%

\section{Results}

%===============================================================================================
\subsection{Nature of the systems}
From the spectroscopic measurements, all of the 17 systems we monitored exhibit spectra that are typical of stars in the RG phase. It is thus easy to track the Doppler shift of the RG spectral lines for each system. All RGs display RV modulations typical of Keplerian orbits with well-resolved amplitudes (12 to 45 km~s$^{-1}$) and periods that match the eclipse periods (Figs.~\ref{fig_ph_rv_sb2} \& \ref{fig_ph_rv_sb1}). Therefore, we can safely deduce that all of the RGs belong to the EB systems we suspected they belonged to, and confirm the nature of the \citet{Gaulme_2013,Gaulme_2014} RG/EB candidates.

The 13 SB2s are the most interesting systems as we can determine the physical properties of each component independently from seismology.  Table~\ref{tab_MR} provides the masses and radii of the RG and companion derived from the eclipse modeling described in \S~\ref{sec_rv}. Fig.~\ref{fig_RG_cmp} shows the masses, radii, and temperatures of each system's components. In agreement with stellar evolution, we find that all companion stars are less evolved and less massive than their RG neighbors. At first glance, the companions are main-sequence K- to F-type stars. Upon closer inspection of Table~\ref{tab_MR}, several of the companion stars do not exactly fall on the main sequence, either because of a larger radius than expected at a given mass and temperature (KICs 4663623, 7037405, 5786154, 9291629), or a temperature that is larger than expected (KIC 7943602). Excessive temperatures or radii could result from RG irradiation and heating by dissipation of tidal energy. The purpose of this paper is not to investigate the evolution of these binary systems, but this will be the subject of future work.

The SB1s also provide some unique astrophysical test cases.  The 1058-day orbit KIC 8054233 is composed of a red clump RG and a MS F-type star on an rather eccentric orbit ($e=0.22$). We were able to extract only three measurements of the companion's RV, which is not enough to get a reliable estimate of its mass. We are not able to efficiently track the companion's spectral lines because the resolution of the ARCES spectrometer does not allow us to clearly disentangle spectra with RV differences less than 10 km\,s$^{-1}$. With its long period, the RVs are separated by only 20 km\,s$^{-1}$ at maximum, which makes it challenging. The three SB1s (KIC 5179609, 5308778, and 8702921) are RGs on the RGB in orbit with M dwarfs of masses $M/M_\sun$ and radii $R/R_\sun$ equal to ($0.6\pm0.01,0.37 \pm0.02$), ($0.64\pm0.03, 0.61 \pm0.02$), and ($0.27\pm0.01,0.28\pm0.01$), respectively.
These three systems's orbital eccentricities are (0.15, 0, 0.1), the RGs show significantly damped oscillation modes, surface activity, and spin-orbit resonances (4:1, 1:1, 5:1) \citep{Gaulme_2014}. It is likely that tidal and radiative interactions have deeply affected their evolutions.

%===============================================================================================
\subsection{Mode suppression in short-period systems is real}

\citet{Gaulme_2014} analyzed the light curves of these systems and concluded that the four with no detectable RG oscillations were indeed RG/EBs. These are four of the six shortest-period binaries in the sample (see Table~\ref{tab_orb} and Table~\ref{tab_MR}). Without any spectra to confirm this, one of their arguments was that the modeled ratio of radii $R\ind{cmp}/R\ind{RG}$ was less than 0.15.  Such a small ratio permits basically two scenarios: a non-oscillating RG with an MS companion, or an exoplanet and an MS star whose pulsations could not be measured in long-cadence data. However, in the case of an exoplanet, none or very shallow secondary eclipses would be observable, while all four of these systems display clear secondary eclipses.

The lack of detectable modes of the RG component can also potentially be explained if it is a younger, less massive star with a $\nu\ind{max}$  larger than the 283\,$\mu$Hz Nyquist frequency of long-cadence data. To test this scenario we apply the inverse asteroseismic scaling relations of \citet{Mosser_2013} to the masses and radii derived from dynamical models (Table~\ref{tab_MR}) to estimate $\nu\ind{max}$, fully aware that the scaling relations, whose accuracy we are testing in this study, can potentially give erroneous results.  Nonetheless, we find that the four RGs have expected $\nu\ind{max}$ values well below the Nyquist frequency.  More precisely, KICs 4569590, 3955867, 9291629, and 7943602 would display $\nu\ind{max}$ of $27\pm1$, $59\pm4$, $61\pm2$ and $74\pm4\ \mu$Hz, respectively.

We therefore confirm that oscillations of RGs in close EBs -- i.e.\ with $(R\ind{RG}+R\ind{cmp})/a \geq 0.16$ -- can be suppressed enough to be undetectable. We stress that here we measure suppression of the entire mode envelope, not the suppression of only $\ell=1$ modes due to internal magnetic fields \citep{Fuller_2015,Stello_2016}. The reasons for this suppression need further investigation.

%-----------------------------------------------------
\begin{figure}[t]
\epsscale{1.2}
\plotone{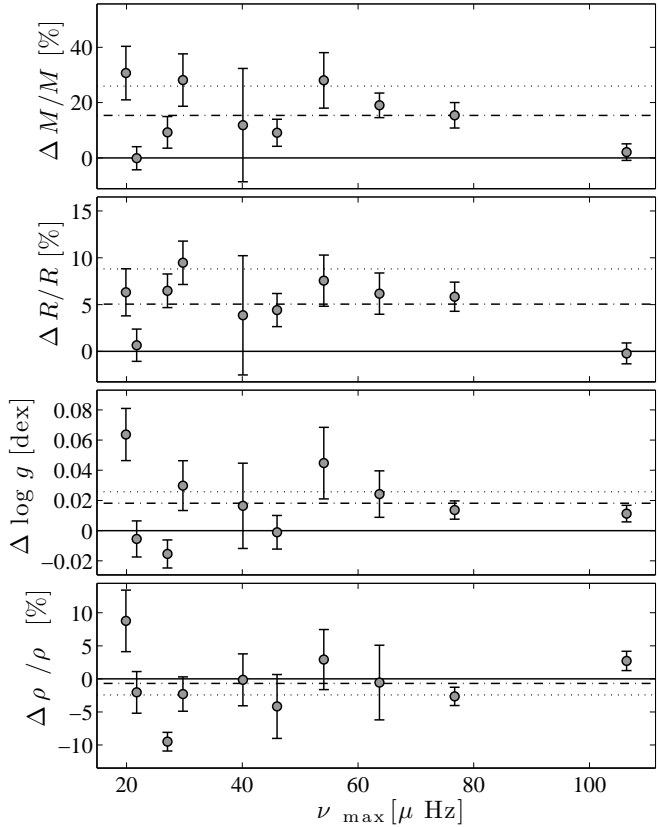}
  \caption{Dynamical modeling versus asteroseismic scaling relations, in the sense of (seismo-RV)/RV, as function of $\nu\ind{max}$. The asteroseismic values are obtained with the \citet{Mosser_2013}'s scaling. The dot-dashed lines indicate the average levels in each panel, and the dotted lines the averages obtained with the ``standard'' scalings including \citet{Kjeldsen_Bedding_1995}'s $\nu\ind{max}$ and $\Delta\nu$. \label{fig_ze_fig}}
\end{figure}

\begin{figure}[t]
\epsscale{1}
\plotone{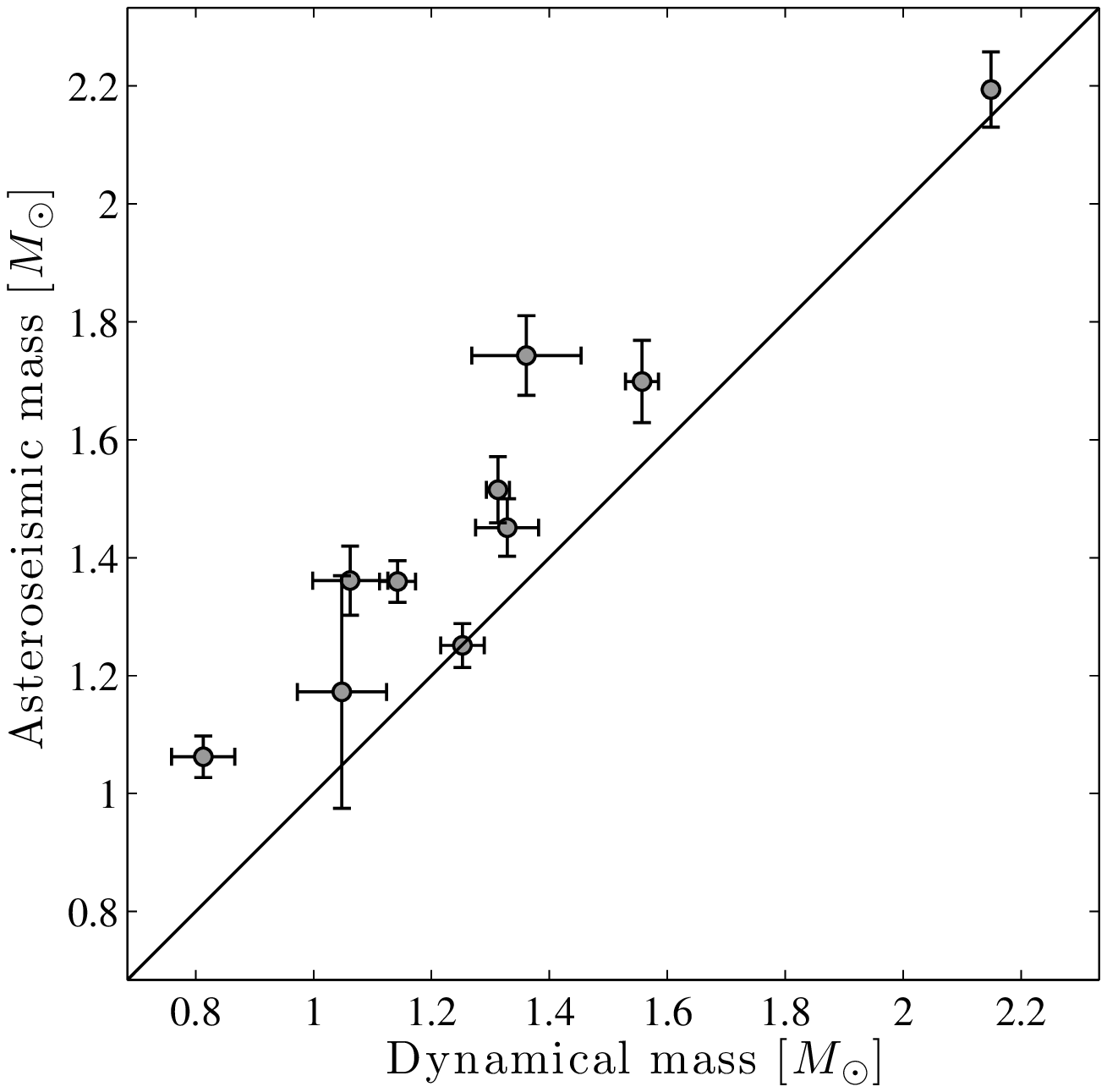}
\plotone{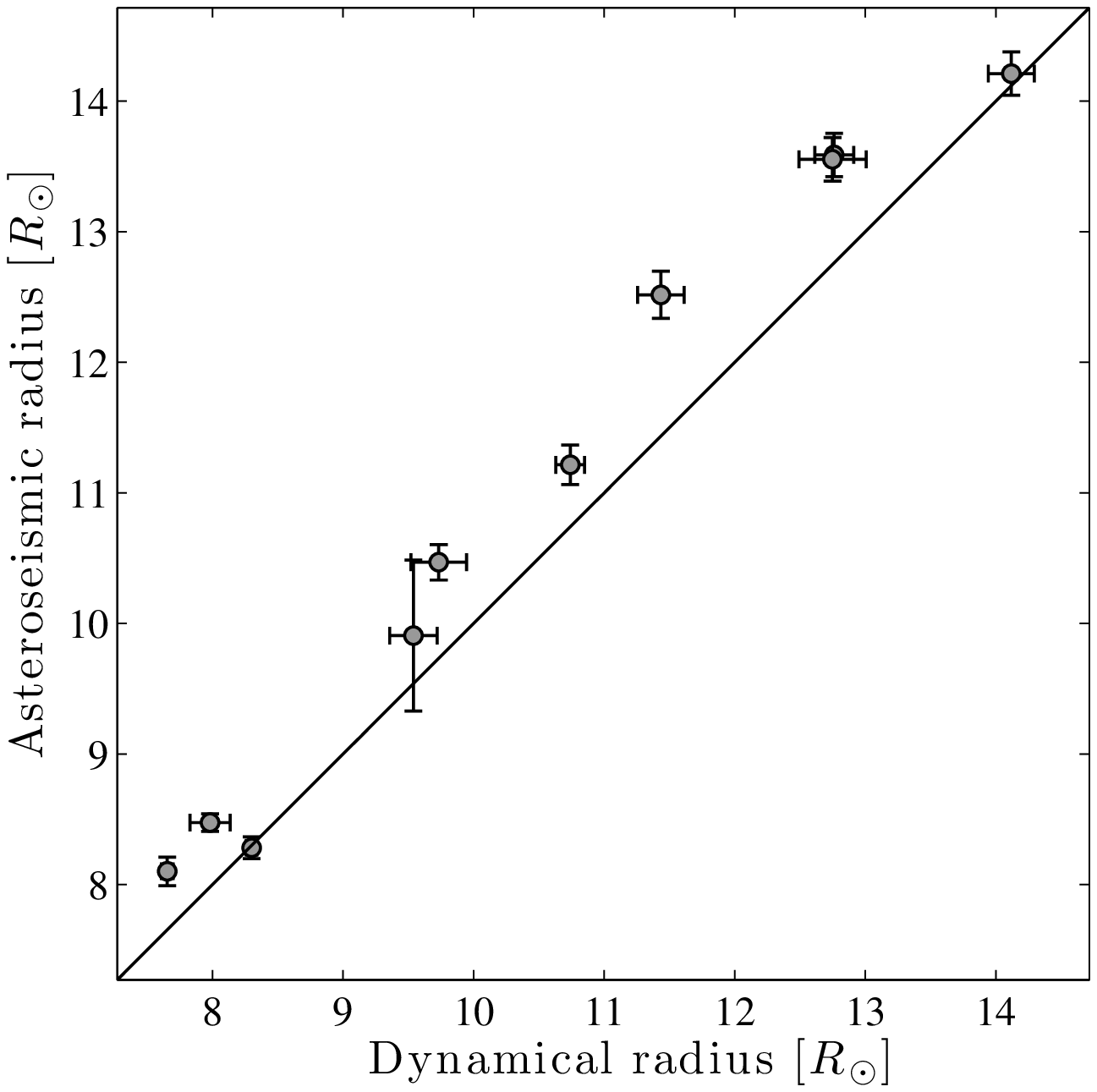}
%\plotone{fig_logg_ast.eps}
  \caption{Masses and radii from \citet{Mosser_2013}'s asteroseismic scaling relations versus those obtained from dynamical modeling. \label{fig_ph_rv_ast}}
\end{figure}
%-----------------------------------------------------

%===============================================================================================
\subsection{asteroseismic scaling relations overestimate mass and radius}

In what follows, we assume that the detailed dynamical modeling gives an accurate representation of the stellar parameters in which we are interested and which we can use to compare to seismic inferences. Figure~\ref{fig_ze_fig} shows the departure of $M/M_\sun$, $R/R_\sun$ $\log g$, and $\bar{\rho}/\bar{\rho_\sun}$ from asteroseismic scaling relations with respect to the dynamical models of the ten SB2 systems with a pulsating RG component. The effective temperatures used as input in the scaling relations are those derived from the visible spectra (Table~\ref{tab_atm}). We also include the estimated parameters of KICs 8410637 and 9246715 from \citet{Frandsen_2013} and \citet{Rawls_2016}.

\begin{figure*}[t]
  \epsscale{0.8}
  \plotone{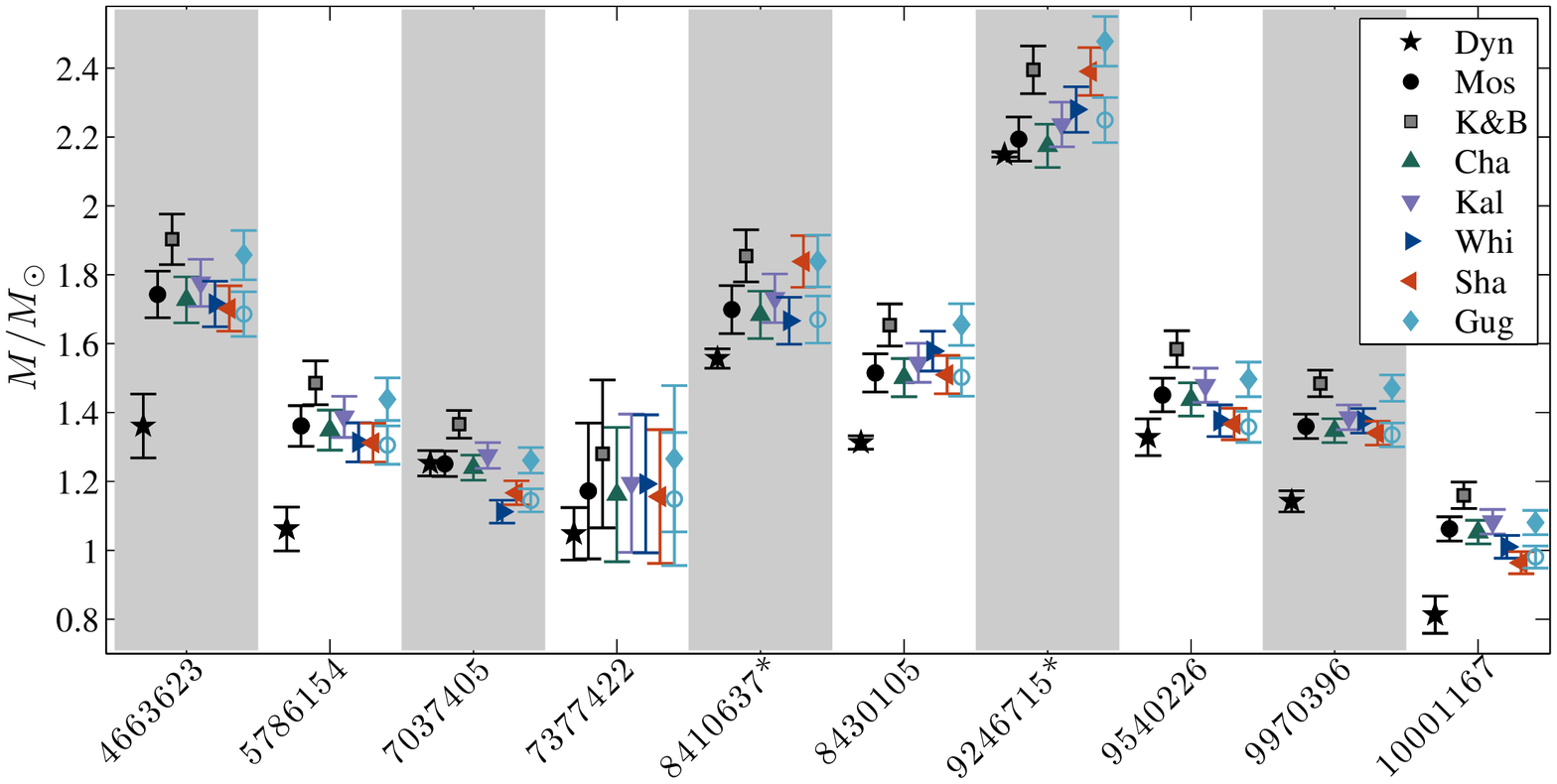}
  \caption{Masses estimates from dynamical modeling (``Dyn'') and seven asteroseismic scaling relations. The abbreviation ``Mos'' stands for \citet{Mosser_2013}, ``K\&B'' \citet{Kjeldsen_Bedding_1995}, ``Cha'' \citet{Chaplin_2011c}, ``Kal'' \citet{Kallinger_2010},  ``Whi'' \citet{White_2011}, ``Sha'' \citet{Sharma_2016}, and ``Gug'' \citet{Guggenberger_2016}. For \citet{Guggenberger_2016}, the diamonds indicate the result with $\nu\ind{max,\odot}=3050\ \mu$Hz and circles with $\nu\ind{max,\odot} = 3150\ \mu$Hz. The asterisks next to the KIC numbers indicate the two likely red clump red giants. A similar trend is observed for radii.}
  \label{fig_scalings}
\end{figure*}

A clear overestimation by seismic scaling is observed for masses and radii for most systems in Figs.~\ref{fig_ze_fig} \& \ref{fig_ph_rv_ast}. On average, \citet{Mosser_2013}'s scaling relation corrections provide masses and radii larger by $15.4\pm10.9\,\%$ and $5.1\pm3.0\,\%$ respectively, while mismatches can reach about 35\,\% and 11\,\% (Table~\ref{tab_MR}). The ``standard'' scaling relations lead to overestimation of masses and radii by $25.9 \pm 11.9\,\%$ and $8.8 \pm 3.1\,\%$ when using $\nu\ind{max,\odot} = 3050\ \mu$Hz \citep{Kjeldsen_Bedding_1995}, $17.6 \pm 11.2\,\%$ and $6.3 \pm 3.0\,\%$ with $\nu\ind{max,\odot} = 3120\ \mu$Hz \citep{Kallinger_2010}, and $14.3 \pm 10.8\,\%$ and $5.4 \pm 3.0\,\%$ with $\nu\ind{max,\odot} = 3150\ \mu$Hz \citep{Chaplin_2011c}.  The temperature-dependent correction of $\Delta\nu$ by \citet{White_2011} leads to masses and radii larger by 13.4\,\% and 4.1\,\% on average. Regarding the study of \citet{Sharma_2016}, adopting their  correction on $\Delta\nu$ provides masses and radii that are overestimated by 13.5\,\% and 4.0\,\% respectively. \citet{Guggenberger_2016}, who worked with \citet{Kjeldsen_Bedding_1995}'s $\nu\ind{max,\odot}$ for their models, are relatively off with respect to the other corrections. However, if we make use of \citet{Chaplin_2011c}'s $\nu\ind{max,\odot}$ in their scaling relations, the output is very similar to the others. Figure~\ref{fig_scalings} summarizes these comparisons for each system.
%these quantities are almost equal to \citep{Kjeldsen_Bedding_1995}'s 

Since both of these quantities (mass, radius) are larger than the dynamical quantities, the discrepancy in surface gravity and mean density (ratios of the two) is not as severe. We measure $\log g$ larger by $0.017\pm0.023$~dex and $0.025\pm0.023$~dex with Mosser and standard scaling relations, whereas densities are lower by $-0.8\pm 4.8\,\%$  and $-2.5 \pm 4.7\,\%$ respectively. There is no clear trend with $\nu\ind{max}$, and the only obvious clump star -- the double-RG system KIC 9246715 -- shows agreement between the measurements.

%While no obvious trend is observed for surface gravity and mean densities, a clear overestimation by seismology  can be observed for masses and radii. For systems with $\nu\ind{max}$ less than $80\ \mu$Hz,  mismatches can reach about 40\,\% in mass and 12\,\% in radius.
%Actually, even very few poins are available, it seems that masses and radii are over estimated by bla bla for $\nu\ind{max} \leq 50\ \mu$Hz.

%%%%%%%%%%%%%%%%%%%%%%%%%%%%%%%%%%%%%%%%%%%%%%%%%%%%%%%%%%%%%%%
%                                                                                   DISCUSSION
%%%%%%%%%%%%%%%%%%%%%%%%%%%%%%%%%%%%%%%%%%%%%%%%%%%%%%%%%%%%%%%
%-----------------------------------------------------

\begin{figure}[t]
\epsscale{1.2}
\plotone{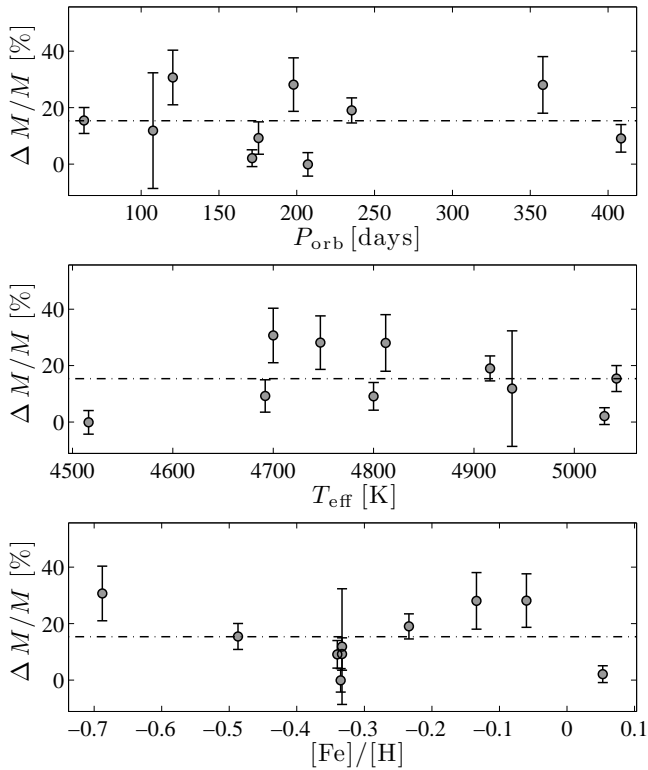}
  \caption{Dynamical modeling versus asteroseismic scaling relations, in the sense of (seismo-RV)/RV, as a function of  $P\ind{orb}$, $T\ind{eff}$, and [Fe/H]. The asteroseismic values are obtained with  \citet{Mosser_2013}'s scaling. The dot-dashed lines indicate the average levels in each panel. \label{fig_dM_notrend}}
\end{figure}
%-----------------------------------------------------

\begin{figure}[t]
\epsscale{1}
\plotone{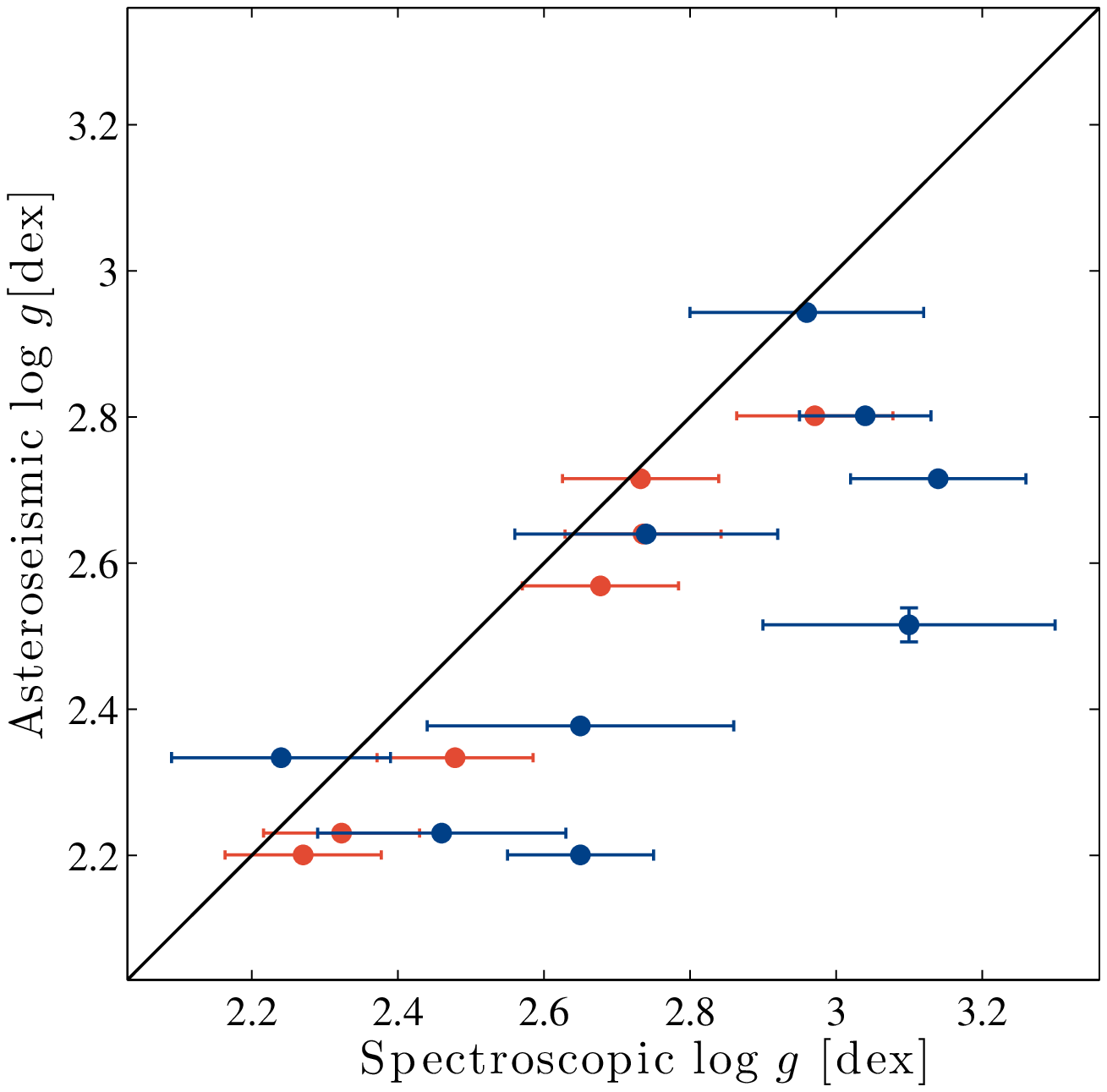}
\plotone{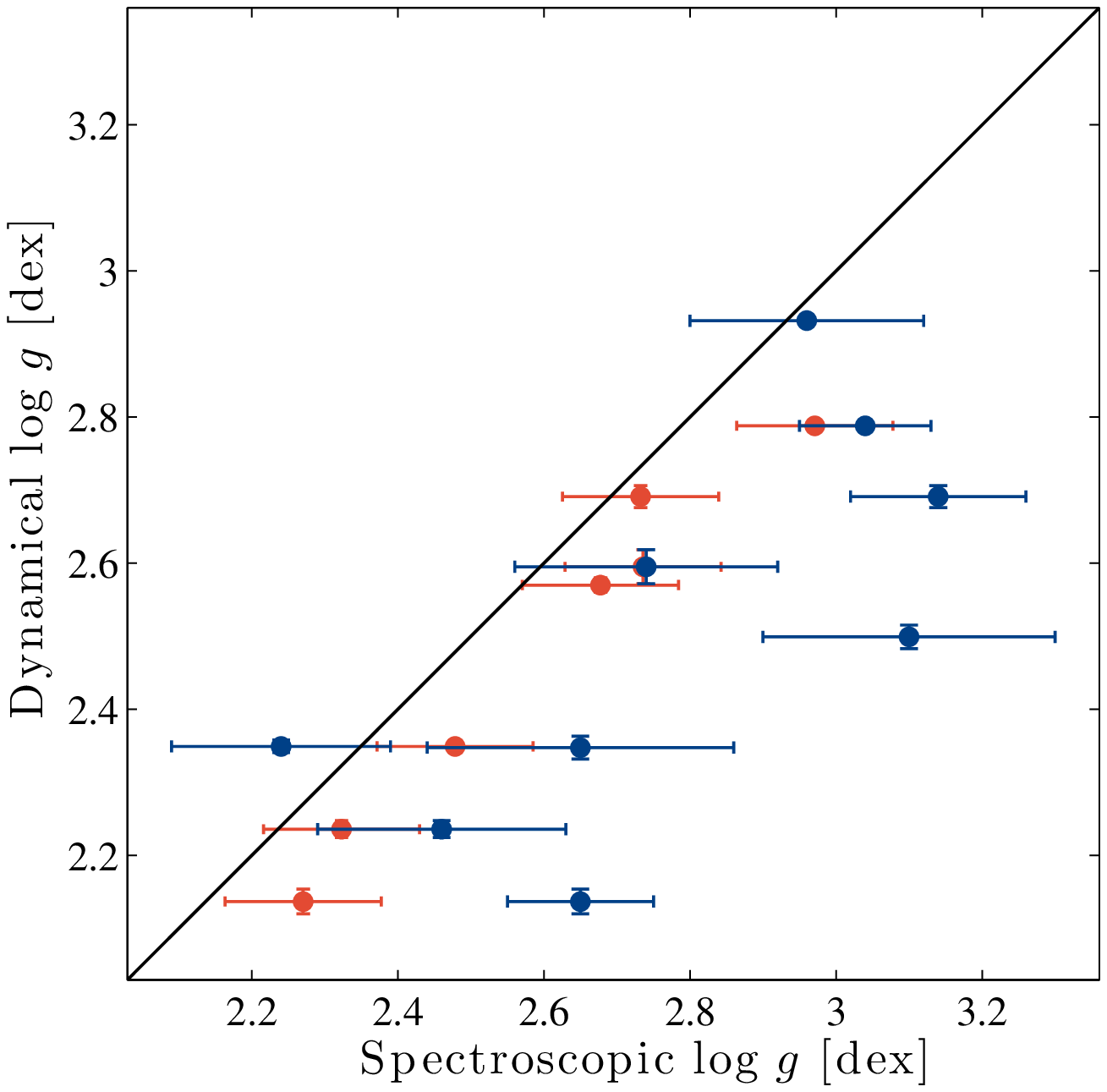}
\caption{Surface gravity $\log g$ from \citet{Mosser_2013}'s asteroseismic scaling relations versus spectrometric estimates with ARCES and APOGEE. Blue symbols indicate the visible ARCES values, and red symbols the IR APOGEE values. \label{fig_logg_rv_ast}}
\end{figure}
%-----------------------------------------------------

\begin{figure}[t]
  \epsscale{1.2}
  \plotone{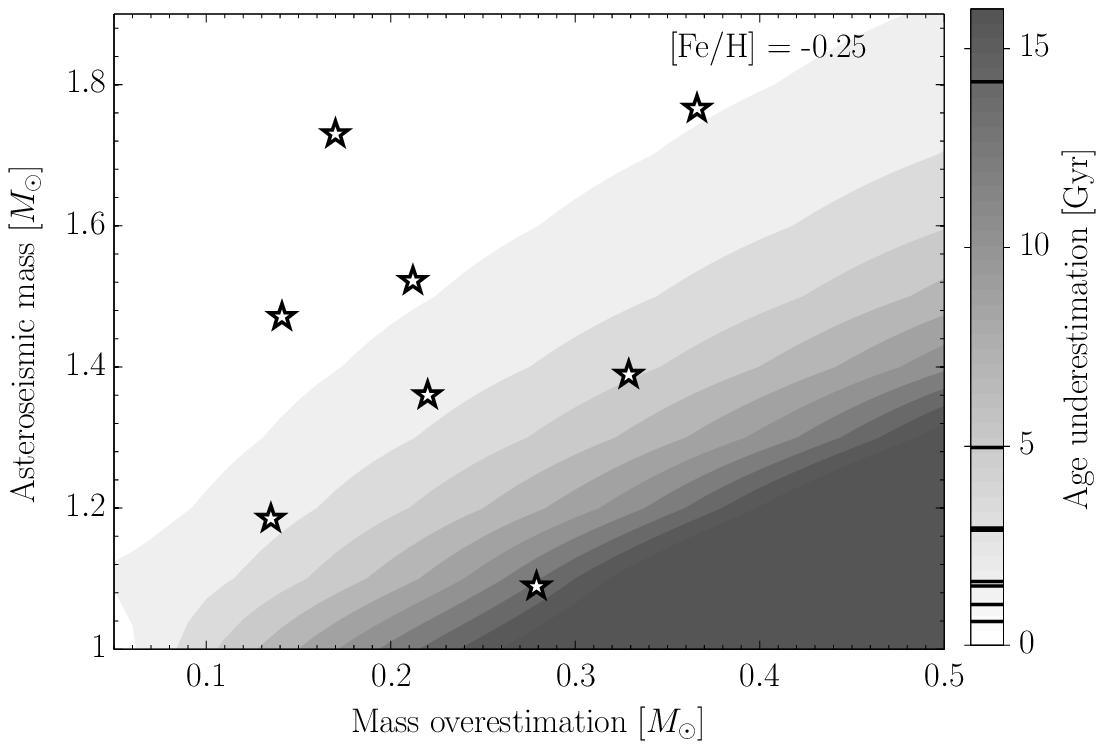}
  \caption{Stellar age overestimation from mass overestimation using isochrones.  The $y$ axis denotes stellar mass measured from seismic scaling relations and the $x$ axis depicts how much larger than the true mass this is. The grey scale indicates the difference in age on the RGB that stellar models would predict for those two initial masses. Also plotted are eight of the stars in this study which have the largest mass discrepancy between seismology and binary modeling. The anticipated age differences for those stars are also indicated on the colorbar with horizontal lines.  The gray scale is clipped at 14\,Gyr. The metallicity of the isochrones is indicated. The isochones were obtained from the BaSTI database \citep{Pietrinferni_2004}.}
  \label{fig_dt_dm}
\end{figure}

\section{Discussion}

The key finding in this study is that asteroseismic masses and radii for a sample of RGs are systematically larger than those obtained through detailed binary modeling.  \citet{Huber_2012} used interferometry of pulsating stars to derive radii and compare to seismic values. While they found a scatter for their 4 red giants similar to what we find, the radii were not systematically larger.
%[[[Compute mean residuals and scatter (rms) to compare?.]]]
\citet{Frandsen_2013} was the first to test the asteroseismic scaling relations with an eclipsing RG star, KIC 8410637. In that analysis the asteroseismic mass and radius were indeed larger than the dynamical ones, the uncertainties were large enough that the comparison was statistically consistent. \citet{Brogaard_2016} has extended that analysis to another eclipsing RG, KIC 9540226 (which we also study) and reach the same conclusion. In our study here, the various shades of the asteroseismic scaling relations we tested lead to similar results. The outliers are those obtained with \citet{Kjeldsen_Bedding_1995}'s solar reference values, regardless of the evolutionary stage, and \citet{Sharma_2016}'s corrections for the two RC giants. The largest influence on  the output from the scaling relations is the choice of  $\nu\ind{max,\odot}$. With a sample composed of ten stars -- mostly RGBs -- masses are overestimated by about 15\,\% and radii by 5\,\%. 

It is important to first understand if there is something specific to these particular stars, i.e., RGs in binary systems, that could  be causing the inconsistencies in these results. The obvious factor here is the potential influence of binarity.  On one hand,  tidal influences on the pulsations could lead to poor measurements of $\Dnu$ and/or $\nu\ind{max}$. Indeed, as discussed earlier, four RGs in short periods have undetectable modes - a fraction much larger than one would expect. These happen to be in short-period orbits. Others have lower pulsation amplitudes than expected. On the other hand, tides could also distort the shape of the stars, leading to models that provide inaccurate radii or other orbital parameters. Any such scenarios should show a trend in the mass and radius disagreement with orbital period; however, the results in Fig. \ref{fig_dM_notrend} (top panel) do not support this. Even the longest-period systems have some of the largest mismatches in mass and radius. Thus,  even if binarity does significantly suppress mode amplitudes, the frequency information appears to be largely unaffected (but with a loss of precision). Furthermore,  most of the systems do not show appreciable  phase effects (e.g., ellipsoidal variations) out of eclipse in the time series, and moreover, such effects would not alter the mass estimation, which is mainly obtained from the radial-velocity data. As these are detached systems, we do not consider the effects of mass transfer between stars. Note that there is no dependence of the mass or radius overestimation as a function of $T\ind{eff}$ or [Fe/H] (Fig.~\ref{fig_dM_notrend}).

Another potential culprit in the scaling relations is the effective temperature. Overestimated temperatures can indeed lead to larger values of mass and radius from the scaling laws even though the functional dependence on $T_{\rm eff}$  is rather weak (to the 3/2 and 1/2 power for mass and radius, respectively). The temperatures we use are derived from visible spectra that do contain the flux from the companion star, although the companions are relatively faint.  Figure~\ref{fig_arces_apogee} shows that the visible temperatures we use in the scaling relations are often larger than the APOGEE-derived ones. Their median difference is 101\,K. Decreasing the temperatures artificially by 100\,K shifts the asteroseismic masses lower by 3.1\,\% and radii by 1.0\,\%, which does not change our conclusions.

The other quantities, $\nu\ind{max}$ and $\Delta\nu$, are almost always straightforward to measure and have rather small uncertainties. Figure~\ref{fig_ph_rv_ast} shows a scatter plot of the seismic and binary masses and radii. In a few cases, the large errors on the seismic quantities are due to suppressed modes, making $\nu\ind{max}$ more difficult to measure precisely. Still, the systematic remains. Furthermore, we find that the spectroscopic gravity is consistently larger than the seismic gravity, which has also been observed in the large APOKASC survey \citep[e.g.,][]{Pinsonneault_2014,Holtzman_2015}. Figure~\ref{fig_logg_rv_ast} shows that this is also the case when compared to the surface gravities obtained from the light-curve modeling masses and radii.

If we exclude KIC 10001167 in Fig.~\ref{fig_ze_fig}, which has the lowest $\nu\ind{max}$ and metallicity, the asteroseismic density is systematically underestimated. Since the density relies exclusively on the $\Delta\nu$ measurement, this suggests that not using the universal pattern and measuring the individual frequencies could perhaps yield a different large separation.

One of the most important consequences of these results is that mass overestimation leads to age underestimation. Figure~\ref{fig_dt_dm} shows this quantitatively for a theoretical collection of stars. Consider RGs whose masses are determined asteroseismically (along the $y$ axis). For that given mass, the $x$ axis is the amount the mass might potentially be overestimated, and the grey scale exhibits the difference in age between stars of the higher and lower mass on the RGB. The data were obtained from BaSTI (``standard scaled solar'') isochrones \citep{Pietrinferni_2004}. Also plotted are eight of the stars from our sample whose mass is overestimated by at least $0.1M_\odot$. While this is just an approximate demonstration since we fix the metallicity of the isochrones and do not take into account uncertainties on the masses, for example, it does illustrate the significant errors one could make in determining the age of red giants, particularly for inherently low-mass stars. For example, seismology would predict KIC 10001167 to be over 12 billion years younger than it would be if its mass were what is determined through binary modeling. More metal-rich isochrones would increase the age overestimation, while more metal-poor ones would decrease it.

If this is a systematic effect for all pulsating red giants, it could have consequences for other current work. For example, the recently discovered $\alpha$-enhanced red giants \citep{Chiappini_2015,Martig_2015} that appear young due to  their rather large (seismic) masses, could indeed be less massive with ages that are more in line with other $\alpha$-enhanced stars.\footnote{There are other reasons to believe these stars are indeed young, however, such as their location in the galaxy.}  The interpretation of large galaxy surveys using asteroseismic data in the context of galaxy population modeling is also an area where these results could be important. For example, \citet{Sharma_2016} concluded that the galaxy population model overpredicts the number of low-mass stars when compared to seismic inferences of red giants. An alternative interpretation given our findings is that seismology may be overpredicting the number of high-mass stars.

%RG and companions (nature of systems, everything is consistent, larger rg and larger companion, eff. temps)

%It's much ``cheaper'' to do asteroseismology than spectroscopy, which gives you a mass and radius (actually, it would be interesting to do this economics. Take Kepler and its cost, and its K1 mission, a few hundred thousand stars. How much does that cost per mass/radius compared to doing it from the ground?)

%--> Excellent idea! Benoit tells me we have the spectra of 17000 RG et 650 MS with K1. Rafa thinks it's 20000RG et ~550 MS. We can roughly consider 20,000 M,R with K1. We should also compute how many 1/2 nights did we observe. The 3.5m cost 1.5M$ per year, i.e. XX per half night etc...

\section{Conclusions}

We have identified and studied a key set of red giant stars in eclipsing binaries that allow for independent methods to obtain masses and radii. By choosing the masses and radii obtained from the binary modeling as the ground truth, we find that (all) seismic scaling relations overestimate both quantities. 

Our measurements  will be of tremendous use in detailed modeling of these red giants that may yield insights into how the scaling laws break down away from the asymptotic limit. In any case, a sense of caution is needed when applying the scaling laws to large samples of giants if, for example, a high accuracy on ages is needed. Even though it may be likely that a simple empirical recalibration of the scaling laws for evolved stars can be applied, as many recent studies have attempted, a more satisfactory understanding is certainly desired.

It is also critical to increase the sample size. We have recently found 16 more RG/EB candidates (Gaulme et al., in prep) which will be promising systems to verify the findings in this work. Among those 16, 10 display oscillations, of which six are SB2s. We have started monitoring their RVs in early 2016, both with the \'echelle spectrographs ARCES at APO and HERMES of the Mercator telescope at La Palma Observatory.

%%%%%%%%%%%%%%%%%%%%%%%%%%%%%%%%%%%%%%%%%%%%%%%%%%%%%%%%%%%%%%%%%%
% TABLE ATM PARAM
\begin{deluxetable*}{l l l l l l l l l}
\tabletypesize{\scriptsize}
\tablecaption{Atmospheric parameters of the red giants from the ARC 3.5-m visible spectra. APOGEE estimates from the DR12 release are indicated in the last three columns when available. Systems are sorted by increasing KIC number.\label{tab_atm}}
\tablewidth{0pt}
\tablehead{
&& \multicolumn{3}{c}{ARCES} &&\multicolumn{3}{c}{APOGEE} \\
\cline{3-5} \cline{7-9} \\
KIC   & $m\ind{Kep}$& $T\ind{eff}$ & $\log g$ &$\left[Fe/H\right]$   && $T\ind{eff}$ & $\log g$&$\left[Fe/H\right]$ \\
          &                      &   [K]             &   [dex]    & [dex]                     & &  [K]              &   [dex]     & [dex]
}
\startdata
3955867 & 13.55 & 4884(83) & 3.2(2) & -0.55(4) & & 4623(91) & 3.0(1) & -0.53(5) \\ 
4569590 & 12.80 & 4706(152) & 2.5(4) & -0.34(9) & & \nodata & \nodata & \nodata \\ 
4663623 & 12.83 & 4812(92) & 2.7(2) & -0.13(6) & & 4803(91) & 2.7(1) & 0.16(4) \\ 
5179609 & 12.78 & 5003(54) & 3.7(2) & 0.22(7) & & 4887(91) & 3.3(1) & 0.45(4) \\ 
5308778 & 11.78 & 4900(44) & 2.5(2) & -0.43(2) & & 5044(91) & 3.3(1) & -0.23(4) \\ 
5786154 & 13.53 & 4747(100) & 2.6(2) & -0.06(6) & & \nodata & \nodata & \nodata \\ 
7037405 & 11.88 & 4516(36) & 2.5(2) & -0.34(1) & & 4542(91) & 2.3(1) & -0.13(6) \\ 
7377422 & 13.56 & 4938(110) & 3.1(2) & -0.33(6) & & \nodata & \nodata & \nodata \\ 
8054233 & 11.78 & 4971(90) & 2.8(2) & -0.15(5) & & \nodata & \nodata & \nodata \\ 
8410637 & 10.77 & \nodata & \nodata & \nodata & & 4699(91) & 2.7(1) & 0.16(3) \\ 
8430105 & 10.42 & 5042(68) & 3.04(9) & -0.49(4) & & 4918(91) & 3.0(1) & -0.43(8) \\ 
8702921 & 11.98 & 5058(86) & 3.3(2) & 0.15(5) & & 4958(91) & 3.3(1) & 0.44(6) \\ 
9246715 & 9.27 & 5030(45) & 3.0(2) & 0.05(2) & & \nodata & \nodata & \nodata \\ 
9291629 & 13.96 & 4713(151) & 3.4(3) & 0.04(6) & & \nodata & \nodata & \nodata \\ 
9540226 & 11.67 & 4692(65) & 2.2(2) & -0.33(4) & & 4662(91) & 2.5(1) & -0.16(8) \\ 
9970396 & 11.45 & 4916(68) & 3.1(1) & -0.23(3) & & 4789(91) & 2.7(1) & -0.18(7) \\ 
10001167 & 10.05 & 4700(66) & 2.6(1) & -0.69(4) & & 4539(91) & 2.3(1) & -0.7(2) \\ 
\enddata
\end{deluxetable*}

%%%%%%%%%%%%%%%%%%%%%%%%%%%%%%%%%%%%%%%%%%%%%%%%%%%%%%%%%%%%%%%%%%
% TABLE ORBITAL PARAM
\newpage
\floattable
\begin{deluxetable*}{l l l l l l l l l l l l l }
\rotate
\tabletypesize{\scriptsize}
\tablecaption{Orbital parameters from dynamical modeling with JKTEBOP. Systems are sorted by decreasing orbital period $P\ind{orb}$. $T\ind{p}$ stands for the time of periastron in Kepler Julian date, $\omega$ the argument of periastron, $e$ the eccentricity, $i$ the orbital plane inclination, $(R_1, T_1, L_1)$ and $(R_2,T_2,L_2)$ the RG and companion's radii, effective temperatures and luminosities. The quantities $K_1, K_2$ are the RV semi-amplitudes and $\gamma$ is the RV offset. The least significant digit in brackets after the value indicates the uncertainty.  \label{tab_orb}}
\tablewidth{0pt}
\tablehead{
KIC & $P\ind{orb}$ & $T\ind{p}^\dagger$ &$\omega$ &$e$& $i$         &$\displaystyle\frac{R_2}{R_1}$& $\displaystyle\frac{R_1+R_2}{a}$& $\displaystyle\left(\frac{T_2}{T_1}\right)^4$& $\displaystyle\frac{L_2}{L_1}$&$K_1$          &$K_2$          &$\gamma$ \\
       & [days]           &    KJD       & [$^\circ$]  &      &[$^\circ$] &         [\%]                &          [\%]                &                                                &            [\%]          &[km s$^{-1}$]&[km s$^{-1}$]&[km s$^{-1}$] \\
}
\startdata
8054233 & 1058.16(2) & -27.69(2) & 302.22(6) & 0.2718(4) & 89.45(1) & 10.83(6) & 1.924(7) & 2.65(4) & 3.453(5) & 12.3(2) & \nodata & -8.68(5)\\ 
4663623 & 358.0900(3) & 129.73(2) & 270.25(2) & 0.43(1) & 88.562(6) & 18.7(3) & 3.91(5) & 4.0(1) & 14.400(7) & 23.0(7) & 23(1) & -8.3(4)\\ 
9970396 & 235.2985(2) & 142.050(2) & 314(2) & 0.194(7) & 89.5(1) & 14.05(7) & 4.39(8) & 2.83(4) & 5.808(4) & 21.4(2) & 24.0(3) & -15.70(5)\\ 
7037405 & 207.1083(7) & 87.194(9) & 310.9(10) & 0.238(4) & 88.65(9) & 12.73(6) & 8.08(8) & 3.79(4) & 6.663(5) & 23.6(2) & 26.0(3) & -39.21(9)\\ 
5786154 & 197.9180(4) & 170.865(3) & 24.7(4) & 0.3764(9) & 88.74(3) & 13.93(6) & 7.14(3) & 3.57(2) & 7.560(4) & 24.7(4) & 25.7(7) & -6.3(4)\\ 
9540226 & 175.4439(6) & 131.415(9) & 4.1(4) & 0.3880(2) & 90 & 7.72(6) & 7.89(2) & 3.46(3) & 2.110(4) & 23.2(3) & 31.4(5) & -12.51(9)\\ 
10001167 & 120.3903(5) & 110.368(9) & 213(2) & 0.159(3) & 87.5(2) & 7.66(4) & 11.4(2) & 3.01(5) & 1.849(4) & 25.1(1) & 25.9(8) & -103.40(6)\\ 
7377422 & 107.6213(4) & 165.185(7) & 356(1) & 0.4377(5) & 85.82(8) & 9.15(6) & 8.84(8) & 2.36(7) & 1.92(1) & 27.5(2) & 34(1) & -56.78(8)\\ 
8430105 & 63.32713(3) & 152.7374(4) & 349.3(2) & 0.2564(2) & 89.01(10) & 10.06(2) & 9.78(3) & 1.716(8) & 1.720(3) & 27.5(2) & 43.7(3) & 16.29(7)\\ 
5179609 & 43.931080(2) & 137.3016(3) & 124.1(1) & 0.150(1) & 86.47(5) & 10.57(2) & 6.92(1) & 2.0(4) & 2.4(1) & 25.0(4) & \nodata & -21.4(2)\\ 
4569590 & 41.3710(1) & 164.286(5) & 261(4) & 0.004(1) & 88.6(6) & 6.85(4) & 21.7(1) & 3.54(7) & 1.615(6) & 34.1(5) & 51(1) & 24.6(1)\\ 
5308778 & 40.5661(3) & 137.281(5) & 272(3) & 0.006(5) & 82.6(2) & 6.02(3) & 17.4(3) & 0.66(2) & 0.222(2) & 23.8(1) & \nodata & 17.406(9)\\ 
3955867 & 33.65685(7) & 160.104(3) & 254(2) & 0.019(2) & 88.0(1) & 11.38(5) & 15.98(6) & 2.79(3) & 3.923(8) & 37.9(2) & 45(1) & 14.82(4)\\ 
9291629 & 20.68643(4) & 154.288(1) & 265(2) & 0.007(2) & 84.10(3) & 23.23(4) & 23.65(5) & 2.70(1) & 15.10(2) & 50.2(2) & 51.2(5) & -30.97(5)\\ 
8702921 & 19.38446(2) & 141.0929(7) & 173(3) & 0.0964(8) & 86.2(3) & 5.34(2) & 15.6(3) & 0.076(2) & 0.0227(6) & 14.0(3) & \nodata & -10.28(9)\\ 
7943602 & 14.69199(4) & 142.542(3) & 103(5) & 0.001(3) & 81.55(7) & 12.63(6) & 24.40(9) & 2.54(3) & 3.48(2) & 46.0(8) & 58(3) & -185.0(1)\\ \enddata
\tablecomments{$^\dagger$ Kepler Julian dates KJD are related to barycentric Julian dates BJD by: KJD = BJD - 2454833 days.}
\tablecomments{$^\star$ As regards 9540226, we fixed the inclination at $90^\circ$ as JKTEBOP would not converge properly and as its inclination is almost $90^\circ$, as the almost vertical ingress and egress of the companion star indicates (Fig. \ref{fig_ph_rv_sb2}).}
\end{deluxetable*}

%%%%%%%%%%%%%%%%%%%%%%%%%%%%%%%%%%%%%%%%%%%%%%%%%%%%%%%%%%%%%%%%%%
% TABLE ASTERO
\newpage
\floattable
\begin{deluxetable}{l l l }
\tabletypesize{\scriptsize}
\tablecaption{Asteroseismic frequencies at maximum amplitude $\nu\ind{max}$ and observed mean large spacings $\Delta\nu\ind{obs}$ of the oscillating RG of our sample. Systems are sorted by increasing KIC number. All $\nu\ind{max}$ were obtained with DIAMONDS but for KIC 7377422, where the low signal-to-noise ratio of the oscillation spectrum prevented the routine from giving an accurate estimate. This specific $\nu\ind{max}$ was fine-tuned with the help of the \'echelle diagram.\label{tab_ast}}
\tablewidth{0pt}
\tablehead{
KIC   & $\nu\ind{max}$& $\Delta\nu\ind{obs}$    \\
          & [$\mu$Hz]       &  [$\mu$Hz]
}
\startdata
4663623 & 54.09 $\pm $ 0.24 & 5.212 $\pm $ 0.019 \\ 
5179609 & 321.84 $\pm $ 1.00 & 22.210 $\pm $ 0.050 \\ 
5308778 & 48.47 $\pm $ 1.10 & 5.050 $\pm $ 0.050 \\ 
5786154 & 29.75 $\pm $ 0.16 & 3.523 $\pm $ 0.014 \\ 
7037405 & 21.75 $\pm $ 0.14 & 2.792 $\pm $ 0.012 \\ 
7377422 & 40.10 $\pm $ 2.10 & 4.643 $\pm $ 0.052 \\ 
8054233 & 46.49 $\pm $ 0.33 & 4.810 $\pm $ 0.015 \\ 
8410637 & 46.00 $\pm $ 0.19 & 4.641 $\pm $ 0.017 \\ 
8430105 & 76.70 $\pm $ 0.57 & 7.138 $\pm $ 0.031 \\ 
8702921 & 195.57 $\pm $ 0.47 & 14.070 $\pm $ 0.010 \\ 
9246715 & 106.40 $\pm $ 0.80 & 8.310 $\pm $ 0.020 \\ 
9540226 & 27.07 $\pm $ 0.15 & 3.216 $\pm $ 0.013 \\ 
9970396 & 63.70 $\pm $ 0.16 & 6.320 $\pm $ 0.010 \\ 
10001167 & 19.90 $\pm $ 0.09 & 2.762 $\pm $ 0.012 \\ 
\enddata
\end{deluxetable}

%%%%%%%%%%%%%%%%%%%%%%%%%%%%%%%%%%%%%%%%%%%%%%%%%%%%%%%%%%%%%%%%%%
% TABLE M, R, LOGG, RHO, ETC
\newpage
\floattable
\begin{deluxetable*}{l l l l l l l l l l l l l l }
\rotate
\tabletypesize{\scriptsize}
\tablecaption{Stellar physical parameters from dynamical modeling (subscripts ``rv'') and asteroseismic scaling relations (subscripts ``ast''). The parameters $M, R, \log g$, and $\bar{\rho}$ refer to stellar masses, radii, surface gravities and mean densities, and $T\ind{eff}$ effective temperatures. Systems are sorted by decreasing orbital period.\label{tab_MR}}
\tablewidth{0pt}
\tablehead{
%&\multicolumn{5}{c}{RG atm. param.\tablenotemark{a}}
%&
%\multicolumn{12}{c}{dynamical model\tablenotemark{b}} \\
%\cline{2-5} \cline{7-19} \\
& \multicolumn{9}{c}{Red Giant} &&\multicolumn{3}{c}{Companion} \\
\cline{2-10} \cline{12-14} \\
%\hline
%KIC  SBtype Mast Mbin R... logg... rho... Teff_RG Mcmp R_cmp ... Teff_cmp
KIC   &  $M\ind{rv}$  & $M\ind{ast}$  &$R\ind{rv}$  & $R\ind{ast}$  &  $\log g\ind{rv}$ & $\log g\ind{ast}$ & $\bar{\rho}\ind{rv}$ &  $\bar{\rho}\ind{ast}$ &  $T\ind{eff}$ & & $M$ & $R$ & $T\ind{eff}$ \\
 &   [$M_\odot$]  & [$M_\odot$]  & [$R_\odot$]&   [$R_\odot$]&  [dex]   & [dex]   & [$10^{-3}\bar{\rho}_\odot $]  &  [$10^{-3}\bar{\rho}_\odot $]  &[K]  & &[$M_\odot$]& [$R_\odot$]& [K] \\
}
\startdata
\multicolumn{14}{c}{Double-line Spectroscopic Binaries (SB2)}\\ 
\hline 
8410637 & 1.56(3) & 1.70(7) & 10.7(1) & 11.2(2) & 2.57(1) & 2.569(5) & 1.26(6) & 1.205(9) & 4800(100) &  & 1.32(2) & 1.57(3) & 6490(160) \\ 
4663623 & 1.36(9) & 1.74(7) & 9.7(2) & 10.5(1) & 2.60(2) & 2.640(5) & 1.48(6) & 1.52(1) & 4812(92) &  & 1.34(7) & 1.82(6) & 6808(140) \\ 
9970396 & 1.14(3) & 1.36(4) & 8.0(2) & 8.47(7) & 2.69(2) & 2.716(3) & 2.2(1) & 2.234(7) & 4916(68) &  & 1.02(2) & 1.12(2) & 6378(91) \\ 
7037405 & 1.25(4) & 1.25(4) & 14.1(2) & 14.2(2) & 2.24(1) & 2.230(3) & 0.45(1) & 0.436(4) & 4516(36) &  & 1.14(2) & 1.80(2) & 6303(53) \\ 
5786154 & 1.06(6) & 1.36(6) & 11.4(2) & 12.5(2) & 2.35(2) & 2.377(5) & 0.71(2) & 0.694(6) & 4747(100) &  & 1.02(4) & 1.59(3) & 6527(138) \\ 
9540226 & 1.33(5) & 1.45(5) & 12.8(1) & 13.6(2) & 2.349(8) & 2.334(4) & 0.639(8) & 0.578(5) & 4692(65) &  & 0.98(3) & 0.99(1) & 6399(90) \\ 
9246715 & 2.149(7) & 2.19(6) & 8.30(4) & 8.28(8) & 2.932(4) & 2.943(4) & 3.76(5) & 3.86(2) & 5030(45) &  & 2.171(7) & 8.37(5) & 4990(90) \\ 
10001167 & 0.81(5) & 1.06(4) & 12.7(3) & 13.6(2) & 2.14(2) & 2.200(4) & 0.39(2) & 0.427(4) & 4700(66) &  & 0.79(3) & 0.98(2) & 6191(91) \\ 
7377422 & 1.05(8) & 1.2(2) & 9.5(2) & 9.9(6) & 2.50(2) & 2.52(2) & 1.21(4) & 1.21(3) & 4938(110) &  & 0.85(3) & 0.87(2) & 6120(143) \\ 
8430105 & 1.31(2) & 1.52(6) & 7.65(5) & 8.1(1) & 2.788(4) & 2.802(4) & 2.93(3) & 2.85(3) & 5042(68) &  & 0.83(1) & 0.770(5) & 5771(78) \\ 
4569590 & 1.56(10) & \nodata & 14.1(2) & \nodata & 2.33(1) & \nodata & 0.56(1) & \nodata & 4706(152) &  & 1.05(4) & 0.96(2) & 6456(211) \\ 
3955867 & 1.10(6) & \nodata & 7.9(1) & \nodata & 2.68(1) & \nodata & 2.19(4) & \nodata & 4884(83) &  & 0.92(3) & 0.90(1) & 6312(108) \\ 
9291629 & 1.14(3) & \nodata & 7.99(5) & \nodata & 2.691(5) & \nodata & 2.24(2) & \nodata & 4713(151) &  & 1.12(2) & 1.86(1) & 6041(194) \\ 
7943602 & 1.0(1) & \nodata & 6.6(2) & \nodata & 2.79(2) & \nodata & 3.40(9) & \nodata & 5096(100) &  & 0.78(5) & 0.83(2) & 6431(128) \\ 
\hline 
\multicolumn{14}{c}{Single-line Spectroscopic Binaries (SB1)}\\ 
\hline 
8054233 & \nodata & 1.60(6) & \nodata & 10.7(1) & \nodata & 2.581(5) & \nodata & 1.294(8) & 4971(90) &  & 1.10(4) & 1.16(2) & 6344(117)  \\ 
5179609 & \nodata & 1.18(3) & \nodata & 3.50(3) & \nodata & 3.423(3) & \nodata & 27.6(1) & 5003(54) &  & 0.60(1) & 0.370(3) & 5950(304)  \\ 
5308778 & \nodata & 1.5(1) & \nodata & 10.1(3) & \nodata & 2.60(1) & \nodata & 1.43(3) & 4900(44) &  & 0.64(3) & 0.61(2) & 4416(52)  \\ 
8702921 & \nodata & 1.67(5) & \nodata & 5.32(5) & \nodata & 3.209(4) & \nodata & 11.07(2) & 5058(86) &  & 0.274(9) & 0.284(3) & 2654(49)  \\ 
\enddata
\tablenotetext{a}{For SB1 systems, the parameters of the companion stars are deduced by combining asteroseismic masses and radii of the RG with the mass function obtained from light curve and radial velocity modeling. }
\tablecomments{The  dagger symbols $^\dagger$ indicate that the dynamical values of KICs 8410637 and 9246715 are taken from \citet{Frandsen_2013} and \citet{Rawls_2016} respectively.}
\end{deluxetable*}

%%%%%%%%%%%%%%%%%%%%%%%%%%%%%%%%%%%%%%%%%%%%%%%%%%%%%%%%%%%%%%%%%%
%         IV. APPENDIX: RV DATA
%%%%%%%%%%%%%%%%%%%%%%%%%%%%%%%%%%%%%%%%%%%%%%%%%%%%%%%%%%%%%%%%%%
\newpage
\clearpage
\appendix
Complete set of radial velocities we present in this paper.

% TABLE PHOTO DYN
\begin{deluxetable}{l l l }
\tabletypesize{\scriptsize}
\tablecaption{Radial velocities (part I). \label{tab_4}}
\tablewidth{0pt}
\tablehead{
Date & RV1                 & RV2                \\
KJD  & [km s$^{-1}$]   & [km s$^{-1}$]  \\
}
\startdata
 \multicolumn{3}{c}{KIC 3955867} \\
 \hline
1704.7441 & -16.20(5) & 51.9(5)\\
1741.7754 & 1.85(5) & 22.7(2)\\
1766.5527 & -19.15(5) & 55.8(3)\\
1936.9669 & -22.96(5) & 61.4(3)\\
1958.9400 & 37.85(6) & -8.9(5)\\
1990.8794 & 45.86(5) & -19.9(6)\\
2032.6426 & -5.46(5) & 40.5(7)\\
2111.7190 & 0.43(5) & 31.4(4)\\
2113.5592 & 14.49(5) & \nodata\\
2121.6013 & 52.47(4) & -29.9(4)\\
2125.7532 & 43.68(5) & -21.7(2)\\
2126.7446 & 37.69(5) & -15.4(5)\\
2286.9775 & 45.42(5) & -21.6(3)\\
2315.8251 & 15.98(4) & \nodata\\
2315.8500 & 16.67(4) & \nodata\\
\hline
 \multicolumn{3}{c}{KIC 4569590} \\
 \hline
1741.7931 & 43.87(7) & -10(1)\\
1936.9265 & -5.57(7) & 69.4(8)\\
1958.8371 & 52.25(8) & -12(1)\\
1967.9110 & 9.08(7) & 47(1)\\
1980.8471 & 1.97(6) & 59.0(8)\\
2032.7087 & 51.27(6) & -13.0(7)\\
2069.6493 & 31.99(6) & \nodata\\
2111.7513 & 33.93(7) & \nodata\\
2113.7085 & 42.82(6) & -5(1)\\
2121.6195 & 58.69(6) & -22.9(7)\\
2125.6981 & 47.16(6) & -8.9(4)\\
2126.6056 & 43.86(6) & -4.7(4)\\
2129.5493 & 29.78(7) & \nodata\\
2315.8739 & 20.64(7) & \nodata\\
2315.8975 & 20.38(7) & \nodata\\
2462.6855 & 18.50(7) & \nodata\\
\hline 
\multicolumn{3}{c}{KIC 4663623}\\ 
\hline 
1737.7064 & -7.11(2) & \nodata\\ 
1741.6719 & -8.52(2) & \nodata\\ 
1958.8201 & 16.01(2) & -31.0(3)\\ 
1980.8653 & 12.99(2) & -30.2(3)\\ 
2032.6816 & 6.25(2) & -21(2)\\ 
2069.6872 & -4.53(2) & -13(4)\\ 
2113.6769 & -11.13(4) & -1(4)\\ 
2121.7263 & -10.87(5) & -1(3)\\ 
2126.6409 & -13.93(2) & 1(7)\\ 
2462.6391 & -11.27(2) & \nodata\\ 
2475.7933 & -12.71(2) & \nodata\\ 
2487.5708 & -12.22(3) & 1(6)\\ 
2639.9061 & -5.46(2) & \nodata\\ 
2670.8693 & 13.17(2) & -32.2(3)\\ 
2674.8132 & 13.00(2) & -31.8(3)\\ 
2685.9605 & 14.79(2) & -30.9(3)\\ 
2736.8396 & 6.57(2) & -21.5(3)\\ 
2780.6121 & -2.25(3) & \nodata\\ 
2780.6286 & -2.13(2) & \nodata\\ 
\hline
\multicolumn{3}{c}{KIC 5179609}\\
\hline
1257.8837 & -30.77(3) & \nodata\\
1272.7406 & -33.86(3) & \nodata\\
1341.9043 & -25.34(3) & \nodata\\
1569.8694 & -41.63(3) & \nodata\\
1591.9183 & 7.53(3) & \nodata\\
1611.8624 & -37.44(3) & \nodata\\
1704.6737 & -43.11(3) & \nodata\\
1711.7021 & -36.07(3) & \nodata\\
1737.6376 & -25.79(4) & \nodata\\
1765.5876 & 5.79(3) & \nodata\\
1765.6280 & 5.01(3) & \nodata\\
1939.9000 & 2.11(3) & \nodata\\
1990.8967 & 1.88(3) & \nodata\\
2069.7258 & -5.34(3) & \nodata\\
2111.7697 & -17.07(3) & \nodata\\
2113.7604 & -6.73(3) & \nodata\\
2121.6748 & 3.89(3) & \nodata\\
2125.6580 & -4.89(3) & \nodata\\
2126.5887 & -7.13(3) & \nodata\\
2286.9007 & -18.59(3) & \nodata\\
2286.9223 & -17.95(3) & \nodata\\
2462.5796 & -17.41(3) & \nodata\\
2462.5988 & -18.06(3) & \nodata\\
2506.5914 & -18.11(3) & \nodata\\
\hline
%\enddata
%\end{deluxetable}
%\begin{deluxetable}{l l l }
%\tabletypesize{\scriptsize}
%\tablecaption{Radial velocities (part II). \label{tab_5}}
%\tablewidth{0pt}
%\tablehead{
%Date & RV1                 & RV2                \\
%KJD  & [km s$^{-1}$]   & [km s$^{-1}$]  \\
%}
%\startdata
\multicolumn{3}{c}{KIC 5308778}\\
\hline
1569.9394 & 38.33(4) & \nodata\\
1591.9508 & -0.21(4) & \nodata\\
1611.9306 & 36.32(4) & \nodata\\
1623.8267 & -1.31(4) & \nodata\\
1683.6971 & 34.83(4) & \nodata\\
1737.5909 & 24.54(4) & \nodata\\
1737.7662 & 22.29(4) & \nodata\\
1741.6243 & 9.89(4) & \nodata\\
1765.6147 & 36.30(4) & \nodata\\
1939.9725 & 25.86(4) & \nodata\\
1958.9553 & 5.30(4) & \nodata\\
1990.9345 & -5.07(4) & \nodata\\
2111.7005 & -4.58(4) & \nodata\\
2113.6109 & -6.16(4) & \nodata\\
2121.7445 & 5.25(4) & \nodata\\
2125.5745 & 20.99(4) & \nodata\\
2126.6561 & 23.74(4) & \nodata\\
2286.9375 & 16.88(4) & \nodata\\
\hline
\multicolumn{3}{c}{KIC 5786154}\\
\hline
1272.7870 & -17.16(2) & \nodata\\
1340.8309 & 19.10(2) & \nodata\\
1569.8874 & 6.49(2) & -17.1(4)\\
1591.9356 & -15.03(2) & 2.05(5)\\
1611.8805 & -21.27(2) & 8.5(5)\\
1623.8985 & -22.68(2) & 11.7(3)\\
1704.6241 & -4.94(2) & \nodata\\
1711.6255 & 1.25(2) & -12.3(2)\\
1741.5950 & 26.26(2) & -38.2(3)\\
1765.6455 & 9.62(2) & \nodata\\
1939.8774 & 25.15(2) & -40.6(3)\\
1980.9316 & -9.30(2) & \nodata\\
2032.7296 & -22.03(2) & 10.9(1)\\
2069.7759 & -16.79(2) & \nodata\\
2111.5711 & 4.26(2) & \nodata\\
2113.5789 & 5.45(2) & -18.1(2)\\
2121.7081 & 9.31(2) & -22.5(1)\\
2125.6755 & 13.28(2) & -27.5(3)\\
2126.7248 & 13.64(2) & -29.4(3)\\
2129.5945 & 18.22(2) & -32.6(1)\\
2487.7330 & -8.26(2) & \nodata\\
2506.5403 & 3.05(3) & \nodata\\
\hline
\multicolumn{3}{c}{KIC 7037405}\\
\hline
1623.9163 & -44.44(4) & -31.5(9)\\
1724.7334 & -40.6(1) & \nodata\\
1726.7233 & -38.4(1) & \nodata\\
1727.7210 & -37.4(1) & \nodata\\
1751.6322 & -14.3(1) & \nodata\\
1752.6307 & -13.7(1) & \nodata\\
1924.8930 & -46.4(1) & \nodata\\
1925.9023 & -45.6(1) & \nodata\\
1927.9058 & -43.9(1) & \nodata\\
1928.8729 & -42.9(1) & \nodata\\
1929.8687 & -42.1(1) & \nodata\\
1930.8812 & -41.1(1) & \nodata\\
1936.9120 & -35.26(3) & \nodata\\
1950.8357 & -20.5(1) & \nodata\\
1951.8220 & -19.7(1) & \nodata\\
1952.8254 & -18.8(1) & \nodata\\
1953.7985 & -18.0(1) & \nodata\\
1954.8094 & -17.3(1) & \nodata\\
1955.8431 & -16.5(1) & \nodata\\
1958.7673 & -15.22(4) & -67.6(5)\\
1967.9409 & -11.77(3) & -67.9(8)\\
1979.7451 & -13.7(1) & \nodata\\
1980.8136 & -14.84(3) & -65.0(8)\\
1981.7555 & -14.6(1) & \nodata\\
1982.7855 & -14.9(1) & \nodata\\
1983.7663 & -15.5(1) & \nodata\\
1984.7620 & -15.9(1) & \nodata\\
1985.7646 & -16.3(1) & \nodata\\
1986.7622 & -17.0(1) & \nodata\\
1987.7560 & -17.4(1) & \nodata\\
1990.8100 & -19.47(3) & -59.1(8)\\
2032.6241 & -42.64(3) & \nodata\\
%\enddata
%\end{deluxetable}
%
%\begin{deluxetable}{l l l }
%\tabletypesize{\scriptsize}
%\tablecaption{Radial velocities (part III). \label{tab_4}}
%\tablewidth{0pt}
%\tablehead{
%Date & RV1                 & RV2                \\
%KJD  & [km s$^{-1}$]   & [km s$^{-1}$]  \\
%}
%\startdata
2069.6345 & -55.72(3) & -21.1(9)\\
2111.5547 & -56.78(3) & \nodata\\
2113.5972 & -56.50(3) & -20.5(8)\\
2121.5857 & -53.13(3) & -23.1(8)\\
2125.6083 & -51.09(3) & -24.2(8)\\
2126.6259 & -50.67(3) & -26.4(9)\\
2129.5633 & -48.59(3) & -27.5(6)\\
2286.9600 & -58.05(3) & -19.3(10)\\
2315.9580 & -57.50(3) & -19.3(9)\\
2462.5620 & -50.71(5) & \nodata\\
2475.7807 & -53.66(3) & -23.3(7)\\
2487.6314 & -55.83(3) & -19.4(8)\\
2506.6085 & -59.54(3) & -18.8(8)\\
\hline
\multicolumn{3}{c}{KIC 7377422}\\
\hline
1697.7442 & -61.74(4) & \nodata\\
1711.7231 & -69.59(3) & -39.0(2)\\
1741.7342 & -70.96(3) & -39.3(4)\\
1936.8880 & -72.75(4) & -38.5(4)\\
1958.9203 & -69.58(4) & -41.9(2)\\
1990.8380 & -23.12(4) & -96.2(4)\\
2032.7870 & -68.31(3) & -39.4(4)\\
2069.7564 & -68.27(3) & -44.2(5)\\
2111.6801 & -35.26(3) & -81.5(4)\\
2113.7270 & -41.95(3) & -78.3(5)\\
2121.6539 & -55.06(3) & \nodata\\
2125.6373 & -59.11(3) & \nodata\\
2126.6746 & -61.66(3) & \nodata\\
2315.9421 & -18.81(3) & -102.2(8)\\
\hline
\multicolumn{3}{c}{KIC 7943602}\\
\hline
1704.7023 & -218.7(1) & -143.2(3)\\
1711.7633 & -145.8(1) & -231.7(6)\\
1741.7134 & -154.1(1) & -222.2(4)\\
1939.9178 & -217.7(1) & -144.9(3)\\
1958.8983 & -143.3(2) & -241.8(4)\\
1980.8297 & -223.3(1) & -138.2(5)\\
1990.8576 & -146.5(1) & -234.3(4)\\
2069.7047 & -228.5(1) & -132.5(6)\\
2111.6127 & -201.9(1) & -155.0(7)\\
2113.6475 & -228.4(1) & -131.8(4)\\
2125.5576 & -188.7(1) & \nodata\\
2126.7027 & -210.3(1) & -153.2(4)\\
2462.6563 & -172.0(2) & -198.9(5)\\
\hline
\multicolumn{3}{c}{KIC 8054233}\\
\hline
1737.6793 & -18.11(2) & \nodata\\
1741.6890 & -18.58(2) & \nodata\\
1936.8672 & -17.35(2) & \nodata\\
1939.9853 & -16.26(2) & \nodata\\
1958.8028 & -14.97(2) & \nodata\\
1980.9473 & -13.20(2) & \nodata\\
2032.7695 & -5.82(2) & \nodata\\
2069.7403 & -1.89(2) & \nodata\\
2113.7753 & 1.15(2) & \nodata\\
2121.7568 & 2.09(2) & \nodata\\
2125.7701 & 2.36(2) & \nodata\\
2129.6116 & 4.72(2) & \nodata\\
2462.6225 & -7.37(2) & \nodata\\
2487.5853 & -6.28(2) & \nodata\\
\hline
\multicolumn{3}{c}{KIC 8430105}\\
\hline
1257.8962 & -1.52(2) & 47.1(8)\\
1272.7620 & 0.80(2) & 43.2(9)\\
1332.7746 & -2.76(3) & \nodata\\
1332.9388 & -5.14(2) & \nodata\\
1333.8804 & -2.60(3) & \nodata\\
1340.9038 & 8.43(5) & \nodata\\
1569.8560 & 0.83(2) & 40(2)\\
1591.8649 & 3.74(3) & 38(1)\\
1611.9079 & 50.06(2) & -36(2)\\
1623.9504 & 17.79(2) & \nodata\\
1704.6859 & -3.85(3) & 49(1)\\
1711.6113 & -3.19(2) & 47(1)\\
1737.7522 & 49.07(2) & \nodata\\
1765.6598 & -3.0(4) & 46(2)\\
1936.9847 & 27.30(2) & -0.0(8)\\
1990.9240 & 50.95(2) & -36.3(10)\\
2111.6517 & 40.25(2) & -22.5(9)\\
2113.5453 & 46.46(3) & -30(2)\\
2121.5395 & 44.10(3) & -26(1)\\
2125.6213 & 31.29(2) & -9(1)\\
2126.6890 & 25.64(2) & -3.8(8)\\
2129.5366 & 20.33(3) & \nodata\\
\hline
\multicolumn{3}{c}{KIC 8702921}\\
\hline
1239.8587 & -1.48(7) & \nodata\\
1239.9301 & -1.21(7) & \nodata\\
1257.8566 & 1.54(8) & \nodata\\
1272.7136 & -1.28(7) & \nodata\\
1332.9188 & 1.07(8) & \nodata\\
1333.9232 & 1.62(8) & \nodata\\
1340.8599 & -19.91(8) & \nodata\\
1569.9025 & -3.72(8) & \nodata\\
1591.8791 & -15.99(8) & \nodata\\
1611.9427 & -18.60(7) & \nodata\\
1683.6718 & 1.95(8) & \nodata\\
1737.6059 & -2.25(7) & \nodata\\
1741.6564 & 2.21(8) & \nodata\\
1939.9603 & -12.52(8) & \nodata\\
1958.9676 & -10.18(8) & \nodata\\
1980.9624 & -21.36(7) & \nodata\\
1990.9465 & 0.95(8) & \nodata\\
2111.7367 & -1.58(7) & \nodata\\
2113.6943 & -8.52(7) & \nodata\\
2121.6368 & -17.51(7) & \nodata\\
2125.7331 & -1.82(8) & \nodata\\
2126.7805 & -0.17(8) & \nodata\\
\hline
\multicolumn{3}{c}{KIC 9291629}\\
\hline
1704.7220 & -50.05(9) & -9.5(1)\\
1711.7861 & 20.76(9) & -79.9(1)\\
1741.7565 & -80.69(9) & 18.0(1)\\
1936.9442 & 12.37(8) & -73.4(1)\\
1958.8578 & 17.26(8) & -77.7(1)\\
1980.8990 & 18.53(8) & -80.1(1)\\
2032.6612 & -79.81(8) & 19.0(2)\\
2069.6667 & -53.60(9) & -6.5(1)\\
2111.6369 & -61.39(8) & 1.1(2)\\
2113.6266 & -78.82(8) & 18.1(1)\\
2121.5667 & -3.73(9) & -56.1(1)\\
2121.7727 & -4.00(9) & -62.32(9)\\
2125.7167 & 17.79(8) & -80.76(10)\\
2126.7643 & 9.88(9) & -77.36(9)\\
2315.9212 & -26.15(8) & \nodata\\
\hline
\multicolumn{3}{c}{KIC 9540226}\\
\hline
1257.9142 & -25.15(2) & 7.8(7)\\
1272.7712 & -26.14(2) & 7.7(4)\\
1332.8001 & -7.65(3) & \nodata\\
1332.8900 & -6.28(2) & \nodata\\
1333.9536 & -5.28(2) & \nodata\\
1340.8854 & 1.48(2) & -32.8(8)\\
1569.9639 & -15.21(2) & \nodata\\
1591.9752 & -23.25(2) & 4(3)\\
1623.8139 & -26.23(2) & 6.9(5)\\
1704.6592 & 18.43(2) & -54.1(6)\\
1711.6883 & 20.28(2) & -53.3(6)\\
1737.7419 & -10.24(2) & \nodata\\
\hline
\multicolumn{3}{c}{KIC 9970396}\\
\hline
1569.9516 & 6.63(2) & -43.9(4)\\
1591.9628 & 4.21(2) & \nodata\\
1611.9547 & -3.76(2) & -29.9(6)\\
1623.8389 & -10.27(2) & -23.5(2)\\
1697.7919 & -33.74(2) & 3.2(6)\\
1711.6722 & -34.64(2) & 4.1(5)\\
1737.5785 & -30.52(2) & 1.4(5)\\
1741.5789 & -29.26(2) & 0.0(4)\\
1939.8576 & -35.19(2) & 4.4(3)\\
1958.7833 & -33.76(3) & 6.5(5)\\
1980.8812 & -27.58(2) & -0.3(6)\\
1990.9108 & -22.65(2) & \nodata\\
2032.7486 & 6.01(2) & \nodata\\
2069.7919 & 0.45(2) & -35.8(3)\\
2111.5885 & -17.17(2) & \nodata\\
2113.7456 & -20.41(2) & \nodata\\
2121.6929 & -22.55(2) & -8.9(3)\\
2125.5948 & -22.54(2) & -6.4(4)\\
2125.7828 & -25.14(2) & -10.1(3)\\
2126.5730 & -22.75(2) & -7.9(2)\\
2129.5787 & -24.04(2) & -7.0(2)\\
2475.7632 & -14.59(2) & \nodata\\
2487.7494 & -5.29(2) & -32.7(6)\\
\hline
\multicolumn{3}{c}{KIC 10001167}\\
\hline
1569.8111 & -107.32(2) & \nodata\\
1591.8177 & -83.34(2) & -123.8(8)\\
1611.8137 & -84.09(2) & -123.1(7)\\
1623.8500 & -90.59(2) & -114(2)\\
1648.8167 & -116.85(3) & -87.3(6)\\
1697.7284 & -95.28(2) & -110.1(6)\\
1711.6021 & -83.61(2) & -123.7(7)\\
1737.5673 & -86.59(2) & -122.4(7)\\
1741.6099 & -89.97(3) & -116.0(8)\\
1765.5998 & -113.77(2) & \nodata\\
1958.7502 & -82.01(3) & \nodata\\
1967.9514 & -82.26(2) & -123.0(9)\\
1980.9162 & -87.70(2) & -121(1)\\
1990.8210 & -95.77(2) & -110.1(6)\\
2032.6951 & -130.34(2) & -74.2(9)\\
2111.6633 & -97.47(2) & -111.2(3)\\
2113.6633 & -99.72(2) & \nodata\\
2121.5500 & -106.60(2) & \nodata\\
2125.5424 & -111.15(2) & -97.5(8)\\
2129.6399 & -117.27(2) & \nodata\\
2286.9958 & -113.77(2) & -92(1)
\enddata
\end{deluxetable}

%----------------------------------------------------------------------------------------------------------------------------------------------------

\acknowledgements{
The authors thank P. Beck, S. Hasselquist, D. Stello, S. Frandsen, M. Pinsonneault, and K. Belkacem.  J.J. and P.G. acknowledge support from NASA ADAP grant NNX14AR85G. E.C. is funded by the European Community's Seventh Framework Programme (FP7/2007-2013) under grant agreement N$^\circ$312844 (SPACEINN). CFB and SM acknowledge support from NSF grant AST-1517592. Most of this paper is based on observations obtained with the Apache Point Observatory 3.5-meter telescope, which is owned and operated by the Astrophysical Research Consortium.  Funding for SDSS-III has been provided by the Alfred P. Sloan Foundation, the Participating Institutions, the National Science Foundation, and the U.S. Department of Energy Office of Science. The SDSS-III web site is http://www.sdss3.org/.
SDSS-III is managed by the Astrophysical Research Consortium for the Participating Institutions of the SDSS-III Collaboration including the University of Arizona, the Brazilian Participation Group, Brookhaven National Laboratory, Carnegie Mellon University, University of Florida, the French Participation Group, the German Participation Group, Harvard University, the Instituto de Astrofisica de Canarias, the Michigan State/Notre Dame/JINA Participation Group, Johns Hopkins University, Lawrence Berkeley National Laboratory, Max Planck Institute for Astrophysics, Max Planck Institute for Extraterrestrial Physics, New Mexico State University, New York University, Ohio State University, Pennsylvania State University, University of Portsmouth, Princeton University, the Spanish Participation Group, University of Tokyo, University of Utah, Vanderbilt University, University of Virginia, University of Washington, and Yale University.
}
\software{JKTLD: http://www.astro.keele.ac.uk/jkt/codes/jktld.html}

%%%%%%%%%%%%%%%%%%%%%%%%%%%%%%%%%%%%%%%%%%%%%%%%%%%%%%%%%%

%\bibliographystyle{plainnat}
%\bibliography{bibi}

%\listofchanges

%\input{comments1}

\end{document}